\documentclass[12pt,letterpaper,a4paper]{article}
\usepackage[includeheadfoot,
marginratio={1:1,2:3},
width=525pt,
height=786pt,]{geometry}

\usepackage{amsmath}
\usepackage{amsfonts}
\usepackage{amssymb}
\usepackage{graphicx}
\usepackage{cancel}
\usepackage{empheq}
\usepackage{color}
\usepackage{hyperref}

\usepackage{graphicx}
\graphicspath{{images/}}

\numberwithin{equation}{section}
\usepackage{float}
\restylefloat{table}
\restylefloat{figure}
\usepackage[utf8]{inputenc}
\usepackage{amsmath, amsthm, amssymb, amsfonts}
\usepackage[font=small, labelfont=bf]{caption}
\usepackage{amsmath, amsthm, amssymb, amsfonts}
\usepackage{multirow}
\usepackage{graphicx}
\usepackage{float}
\usepackage[numbers,sort&compress]{natbib}
\usepackage{enumerate}
\usepackage{amsmath}
\usepackage{amsfonts}
\usepackage{amssymb}
\usepackage{graphicx}
\usepackage{cancel}
\usepackage{empheq}
\usepackage{color}
\usepackage{colortbl}
\definecolor{green2}{cmyk}{0, 1, 0.5, 0.3}
\definecolor{green3}{cmyk}{1, 0.75, 1.0, 0.0}

\definecolor{lightgreen}{cmyk}{0.2, 0, 0.2, 0.2}
\definecolor{lightgray}{cmyk}{0.1,0.2,0,0.1}
\definecolor{lightgray2}{cmyk}{0.4,0.4,0,0.8}
\definecolor{black}{cmyk}{1.0,1.0,1.0,1.0}

\usepackage{hyperref}

\restylefloat{table}
\restylefloat{figure}
\usepackage{longtable}
\usepackage{lipsum} 

\usepackage{amsfonts}
\usepackage{amssymb}
\usepackage{comment}
\usepackage{graphicx}
\usepackage{amsmath}
\usepackage{tabu}
\usepackage{dsfont}

\usepackage{amsthm}
\usepackage{slashed}
\usepackage{setspace}
\usepackage[labelformat=simple]{subcaption}

\usepackage{pgfplots}
\usepackage{tikz}
\usepackage{tikz-cd}
\usetikzlibrary{shapes.misc,shadows,fit,decorations.pathmorphing,positioning,trees,decorations.markings,decorations.pathreplacing,calc,shapes,patterns,arrows,positioning,arrows.meta}
\usepackage{xcolor}
\usepackage{pgfplots}
\usepackage{pgfplotstable}
\usepackage{xcolor}
\usepackage{makecell}
\usepackage{diagbox}
\usepackage{environ}
\usepackage[utf8]{inputenc}
\usepackage{amsmath, amsthm, amssymb, amsfonts}
\usepackage{multirow}
\usepackage{graphicx}
\usepackage{float}
\usepackage[numbers,sort&compress]{natbib}
\usepackage{enumerate}
\usepackage{enumitem}
\usepackage[capitalize,sort]{cleveref}
\usepackage{hhline}
\usepackage{mathtools}
\usepackage{longtable}
\usepackage{makecell}
\usepackage{tabularx}
\usepackage{cellspace}
\usepackage{alphalph}
\usepackage{lscape}
\usepackage{slashed}
\usepackage{setspace}
\usepackage{pifont}

\usepackage{float}

\crefname{figure}{Figure}{Figures}
\crefname{table}{Table}{Tables}

\def\be{\begin{equation}}
\def\ee{\end{equation}}
\def\bea{\begin{eqnarray}}
\def\eea{\end{eqnarray}}
\def\bes{\begin{subequations}}
	\def\ees{\end{subequations}}

\def\ov{\overline}

\def\ov{\overline}
\def\1{{\bf 1}}
\def\2{{\bf 2}}
\def\3{{\bf 3}}
\def\4{{\bf 4}}
\def\6{{\bf 6}}

\newcommand{\nn}{\nonumber}

\newcommand{\beq}{\begin{equation}}
\newcommand{\eeq}{\end{equation}}

\def\ov{\overline}

\allowdisplaybreaks[2]
\numberwithin{equation}{section}

\def\be{\begin{equation}}
\def\ee{\end{equation}}
\def\bea{\begin{eqnarray}}
\def\eea{\end{eqnarray}}
\def\bes{\begin{subequations}}
	\def\ees{\end{subequations}}

\usepackage{multicol}
\usepackage{float}
\usepackage{caption}

\usepackage{setspace}

\allowdisplaybreaks[2]
\numberwithin{equation}{section}

\usepackage{amsfonts}
\usepackage{amssymb}
\usepackage{comment}
\usepackage{graphicx}
\usepackage{amsmath}
\usepackage{tabu}
\usepackage{dsfont}
\usepackage{float}
\usepackage{amsthm}
\usepackage{slashed}
\usepackage{setspace}
\usepackage{pgfplots}
\usepackage{tikz}
\usepackage{tikz-cd}
\usetikzlibrary{shapes.misc,shadows,fit,decorations.pathmorphing,positioning,trees,decorations.markings,decorations.pathreplacing,calc,shapes,patterns,arrows,positioning,arrows.meta}
\usepackage{xcolor}
\usepackage{pgfplots}
\usepackage{pgfplotstable}
\usepackage{xcolor}
\usepackage{makecell}
\usepackage{diagbox}
\usepackage{environ}

\usepackage{amsmath, amsthm, amssymb, amsfonts}
\usepackage{amsmath, amsthm, amssymb, amsfonts}
\usepackage{multirow}
\usepackage{graphicx}
\usepackage{float}
\usepackage[numbers,sort&compress]{natbib}
\usepackage{enumerate}
\usepackage{enumitem}
\usepackage{hhline}
\usepackage{mathtools}
\usepackage{longtable}
\usepackage{alphalph}

\usepackage{amsfonts}
\usepackage{amssymb}
\usepackage{comment}
\usepackage{graphicx}
\usepackage{amsmath}
\usepackage{tabu}
\usepackage{dsfont}
\usepackage{float}
\usepackage{amsthm}
\usepackage{slashed}
\usepackage{setspace}

\usepackage{amsmath, amsthm, amssymb, amsfonts}
\usepackage[font=small, labelfont=bf]{caption}
\usepackage{amsmath, amsthm, amssymb, amsfonts}
\usepackage{multirow}
\usepackage{graphicx}
\usepackage{float}
\usepackage[numbers,sort&compress]{natbib}
\usepackage{enumerate}

\usepackage{amsmath}
\usepackage{amsfonts}
\usepackage{amssymb}
\usepackage{graphicx}
\usepackage{cancel}
\usepackage{empheq}
\usepackage{color}
\usepackage{hyperref}

\usepackage{float}
\usepackage{framed}
\restylefloat{table}
\restylefloat{figure}

\usepackage[margin=1cm]{caption}

\def\ov{\overline}

\allowdisplaybreaks[2]
\numberwithin{equation}{section}

\begin{document}
	{\hfill
		\hfill
		arXiv:2606.xxxxx}

	\vspace{1.0cm}
	\begin{center}
		{\Large
        On Calabi-Yau Threefolds For Unified LVS Inflation}
		\vspace{0.4cm}
	\end{center}

\vspace{0.2cm}
\begin{center}
Pramod Shukla$^\dagger$ \footnote{pshukla@jcbose.ac.in}
\end{center}

\vspace{0.2cm}
\begin{center}
{$^\dagger$ Department of Physical Sciences, Bose Institute,\\
Unified Academic Campus, EN 80, Sector V, Bidhannagar, Kolkata 700091, India}\\
\vspace{0.3cm}
\end{center}
\vspace{1cm}

\abstract{
Fibre inflation, Poly-instanton inflation and (Loop) Blow-up inflation are among the most popular K\"ahler moduli based inflationary models realized in the standard LARGE volume scenarios (LVS). In this article, we present a unified framework in which all these three LVS inflationary models can be realized by using (different orientifolds of) a single Calabi-Yau (CY) threefold. In fact, the desired CY threefold needs to have a K3- or ${\mathbb T}^4$-fibration structure along with two diagonal del Pezzo divisors, and a so-called `Wilson' divisor which corresponds to a surface realized as a ${\mathbb P}^1$ fibration over ${\mathbb T}^2$s. For classification purpose, we perform a detailed scan of the CY geometries with $1 \leq h^{1,1}({\rm CY}) \leq 6$ that arise from the triangulation of the four-dimensional reflexive polytopes of the Kreuzer-Skarke database. In this regard, after scanning around 100,000 CY geometries and the corresponding topologies of around a million of toric divisors, we find two CY threefolds satisfying these requirements for $1 \leq h^{1,1}({\rm CY}) \leq 4$, while there are 14 and 45 candidate CY geometries for $h^{1,1}({\rm CY}) = 5$ and $h^{1,1}({\rm CY}) = 6$, respectively. We discuss the extended applications of such CY threefolds for cosmological model building in string theoretic frameworks. 
}

\clearpage

\tableofcontents


\section{Introduction}
\label{sec_intro}

Model building in string cosmology is a delicate task having several layers of challenges, with a number of assumptions considered on the way. This leads to a set of presumptions followed by a set of expectations which are to be tested against reality, say by making attempts to consistently match the experimentally observed values of the various cosmological observables. As a model builder, to begin with, one considers a type IIB compactified on a CY orientifold based setup and starts with a minimal list of requirements such as
\begin{itemize}

\item 
a concrete Calabi-Yau (CY) threefold with desired global properties such as Hodge numbers $(h^{1,1}, h^{2,1})$ in order to fix the number of the K\"ahler moduli and the complex structure moduli in the game.

\item 
further details on the properties of the CY threefolds  such as Euler number $\chi({\rm CY})$ and second Chern class $c_2({\rm CY})$ to quantify the (leading order) $\alpha^\prime$-corrections.

\item 
knowledge of the K\"ahler cone conditions and triple intersection numbers to determine the structure of the volume form such as Swiss-Cheese structure, K3-fibrations etc.

\item a suitable form of the K\"ahler potential and the superpotential to generate an effective $N=1$ four-dimensional scalar potential induced through the F-term contributions. Subsequently, one analyzes such scalar potentials for various purposes such as moduli stabilization and inflationary aspects.

\end{itemize}
Having said that, it is true that the actual demands/constraints for the global model building arise through the substructures of the CY threefolds where one needs the local pieces of information which are harder to compute. For example, the divisor and curve topologies have significant role in generating the effective scalar potential pieces, as various string-loop effects demand specific brane-setting following from a suitably chosen holomorphic involution. All these requirements make the model building quite a delicate process.

The most prominent task for constructing four-dimensional (semi)realistic models using superstring compactifications is about finding the ``right" Calabi Yau (CY) threefold. For more than three decades, an enormous amount of efforts has been aimed towards constructing and classifying CY threefolds which could be suitable for model building \cite{Green:1986ck, Candelas:1987kf, Green:1987cr, Candelas:1993dm, Batyrev:1993oya,Candelas:1994hw,Hosono:1994ax,Kreuzer:2000xy,Gray:2013mja}. These efforts have led to two main CY threefold datasets, namely the Complete Intersection CY threefolds (CICYs) realized as multi-hypersurfaces in the product of projective spaces \cite{Green:1986ck}, and the Toric Hypersurface CY threefolds (THCYs) realized as hypersurfaces in toric varieties \cite{Batyrev:1993oya} arising from appropriate triangulations of the reflexive polytopes classified in the Kreuzer-Skarke (KS) database \cite{Kreuzer:2000xy}. The projective CICY database of ~\cite{Candelas:1987kf,Anderson:2017aux} is sometimes refereed as ``pCICY" database in order to distinguished with the so-called ``generalized" CICYs referred as ``gCICY" \cite{Anderson:2015iia} and the Toric CICY refereed as ``tCICY". A couple of concrete CY threefolds of tCICYs have been presented in \cite{Cicoli:2021dhg} in which the ones with so-called diagonal del-Pezzo surfaces of dP$_5$ type appearing as toric divisors are quite interesting. Further relevant details about the studies and classification of these two CY datasets are as follows: 

\begin{itemize}
    
\item 
A complete list of 7890 CICYs along with their Hodge numbers $(h^{1,1}, h^{2,1})$ have been presented in  \cite{Candelas:1987kf,Green:1987cr} while their fibration structures have been studied in detail in \cite{Gray:2014fla,Anderson:2017aux}. Moreover, some initiatives have been taken towards  MSMS-like local models building \cite{Anderson:2011ns, Anderson:2012yf, Anderson:2013xka},  and moduli stabilization as well \cite{Anderson:2010mh,Anderson:2011cza, Bobkov:2010rf,Carta:2021sms,Carta:2021uwv, Carta:2022web,Carta:2022oex}. Further, a classification of CICY orientifolds with odd $(1,1)$-cohomology sector has been presented in \cite{Carta:2020ohw}, while a thorough analysis and classification of their (toric) divisor topologies are presented in \cite{Carta:2022web}. It is worth mentioning that for pCICYs, most of the topological information, such as triple intersection numbers, second Chern class, Euler characteristics, Mori cone, K\"ahler cone and even the Gopakumar-Vafa invariants \cite{Gopakumar:1998ii,Gopakumar:1998jq}, needed for model building have been computed in a series of works \cite{Anderson:2017aux,Hosono:1994ax,Carta:2021sms}.

\item 
Although computations of some selected topological properties of a few CY geometries of the THCYs database have been initiated in the early nineties \cite{Candelas:1993dm, Candelas:1994hw,Hosono:1994ax}, a systematic exploration of the KS database CYs have been witnessed recently with a huge surge in an efficient computations of topological properties of the CY threefold. This has been possible due to some efficient tools/packages being developed in the meantime. These include, for example, the ``Package for analyzing lattice polytopes" (PALP) \cite{Kreuzer:2002uu}, its updated new offspring version \cite{Braun:2011ik} with \texttt{mori.x} module, SAGE \cite{sagemath}, \texttt{cohomCalg} package  \cite{Blumenhagen:2010pv,Blumenhagen:2011xn}. In fact, there have been tremendous  efforts in the area of computational CY geometries, especially with the advent of the \href{https://cytools.liammcallistergroup.com/about/}{\texttt{CYTools}} which has been proven to be very efficient in triangulating the polytopes, including those corresponding to larger $h^{1,1}$. Moreover, it has been also demonstrated to perform divisor topology computations, and its subsequent phenomenological implications have been initiated in \cite{Braun:2017nhi,Demirtas:2018akl,Demirtas:2020dbm}.

\end{itemize}

\noindent
Our focus for the current work is mainly towards analyzing the CY threefolds of the THCYs database for the purpose of inflationary models developed in the so-called LARGE volume scenarios (LVS) \cite{Balasubramanian:2005zx}. In this regard,  we will heavily use a quite {\it phenomenologist-friendly} subset of the THCYs developed for CYs with $1\leq h^{1,1}(CY) \leq 6$, and referred as Altman-Gray-He-Jejjala-Nelson (AGHJN) dataset \cite{Altman:2014bfa}. This is an amazing data collection equipped with mostly all the necessary model building first-hand information, e.g. GLSM data, SR ideal, Fundamental group, Hodge numbers of the CY threefold, Chern classes, Triple intersection numbers, Mori/K\"ahler cone etc. This database has been subsequently used for constructing models \cite{Altman:2017vzk, Cicoli:2018tcq}. Moreover, as an extension of the previous work in \cite{Gao:2013pra}, this AGHJN-dataset has been updated for orientifolds with non-trivial odd (1,1) cohomology in \cite{Altman:2021pyc}, and further extended for $h^{1,1}({\rm CY}) \geq 7$ in \cite{Altman:2014bfa, Altman:2021pyc, Gao:2021xbs, Crino:2022zjk}.

Finding the `right' CY threefold suitable for realistic string model building remains of the most difficult questions. Nevertheless one can always make a classification for the known lists of CY threefolds to suggest that ``some" class of CY threefold can be more useful for certain purposes as compared to the other ones. In this regard, the structure of compactifying CY has a huge impact on the model building as the internal geometries such as divisor and curve topologies, involutions, brane-setting, tadpole cancelation conditions, K\"ahler cone conditions, etc. play a crucial role in determining the effective scalar potential of the four-dimensional supergravity theory. In the context of inflationary model building within string theoretic framework, one begins with several assumptions (e.g. in the form of Ansatz for the K\"ahler potential ${K}$ and and superpotential $W$) and it is necessary to look for explicit constructions in which those assumptions are truly fulfilled. Therefore, for  constructing a global model of any viable inflationary proposal, the central task reduces into finding the compactifying CY threefold which can result in {\it `appropriate' and `just-enough' corrections}. Both these conditions are necessary; e.g. the holomorphic involution and brane-setting should not only be rich enough to induce appropriate corrections but should also avoid having the unwanted or too many contributions to the effective scalar potential, in order to maintain the flatness of the inflationary potential achieved with the minimal set of terms. In this regard, study of divisor topologies of the compactifying CY threefolds in a systematic way turns out to be a crucial step towards global model building. 

In the current work, our aim is two-fold; first we extend our earlier classification/analysis of the divisor topology of CY threefolds with $1 \leq h^{1,1}({\rm CY}) \leq 5$ (as presented in \cite{Shukla:2022dhz}) by including 84525 more CY geometries corresponding to $h^{1,1}({\rm CY}) = 6$. In this regard, we present a range of divisor topologies including the (diagonal) del-Pezzo divisors, K3-fibration structure of the CY threefold, the so-called Wilson divisors relevant for poly-instanton inflation \cite{Blumenhagen:2012pc}. We also extend the previous results \cite{Shukla:2022dhz} on including the classification of the suitable candidates CY geometries for blow-up inflation, fibre inflation and poly-instanton inflation \cite{Cicoli:2016xae, Cicoli:2017axo, Cicoli:2011ct,Blumenhagen:2012ue,Blumenhagen:2012kz,Conlon:2005jm, Bond:2006nc, Cicoli:2017shd, Bansal:2024uzr} for CY geometries with $h^{1,1}({\rm CY}) = 6$. Subsequently we will aim to find the CY geometries which can satisfy the properties of all the three classes of inflationary models proposed in the LARGE volume scenarios (LVS) \cite{Balasubramanian:2005zx}, leading to what we call a `unified' framework for LVS inflation. Surprisingly we find that the number of such CY geometries which could potentially help in realizing a unified LVS inflation is significantly small as we find only 61 candidates after scanning around 100000 CY geometries having nearly a million of divisor topologies.

The article is organized as follows: Section \ref{sec_setup} presents some basic preliminaries about the LVS moduli stabilization schemes in the type IIB orientifold compactification setups, and the popular inflationary proposal in LVS. Section \ref{sec_topo-classification} includes the detailed classification of the divisor topologies for all the CY geometries having $1\leq h^{1,1}({\rm CY}) \leq 6$. In section \ref{sec_TwoEx-UnifiedLVS} we present the scanning results for the CY geometries suitable for unified LVS inflation and provide two concrete CY examples with relevant details. This section also includes the discussion on the various contributions to the effective scalar potential. Section \ref{sec_unifiedLVS} presents the suitable orientifolds and brane-settings for all the three LVS inflationary models presenting a possible unified framework. Section \ref{sec_conclusions} summarizes the results with outlook, which is followed by an appendix \ref{sec_appendix} that includes a list of 14 candidate CY geometries with $h^{1,1}({\rm CY}) = 5$ suitable for unified LVS inflation.


\section{Preliminaries}
\label{sec_setup}
In this section, we present a brief review about the moduli stabilization in the standard large volume scenario (LVS), and the three popular inflationary models realized in LVS framework. 

\subsection{Moduli stabilization in LVS}

The low energy dynamics of the four-dimensional effective supergravity theory arising from the type IIB superstring compactifications on CY orientifolds can be captured by a holomorphic superpotential ($W$) and a real K\"ahler potential ($K$) and the gauge kinetic function $(g)$. These quantities depend on the various chiral coordinates obtained by complexifying the various moduli with a set of RR axions. We fix the conventions by following definitions of the chiral variables:
\bea
\label{eq:chiral-variables}
& & U^i = v^i - i\, u^i, \qquad S = c_0 + i\, e^{-\phi}, \qquad T_\alpha = c_\alpha - i\, \tau_\alpha~,
\eea
where $\phi$ is the dilaton, $u^i$'s are the complex structure saxions, and $\tau_\alpha$'s are the Einstein frame four-cycle volume moduli defined as $\tau_\alpha = \partial_{t^\alpha} {\cal V} = \frac12 k_{\alpha\beta\gamma} t^\beta t^\gamma$ where $t^\alpha$ is the two-cycle volume and ${\cal V}$ denotes the overall volume of the CY threefold. In addition, the $c_0$ and $c_\alpha$'s are universal RR axion, RR four-form axions respectively while the complex structure axions are denoted as $v^i$. Here the indices $\{i, \alpha\}$ are such that $i \in h^{2,1}_-({\rm CY}/{\cal O})$ while $\alpha \in h^{1,1}_+({\rm CY}/{\cal O})$. Moreover, we assume that $h^{1,1} = h^{1,1}_+$ for simplicity, and hence there are no so-called odd-moduli $G^a$ which are present in our analysis, and we refer the interested readers to \cite{Cicoli:2021tzt}.

The F-term contributions to the scalar potential are computed using the following well known formula,
\bea
\label{eq:V_gen}
& & e^{- {K}} \, V = {K}^{{\cal A} \ov {\cal B}} \, (D_{\cal A} W) \, (D_{\ov {\cal B}} \ov{W}) -3 |W|^2 \equiv V_{\rm cs} + V_{\rm k}\,,
\eea
where:
\bea
\label{eq:VcsVk}
& & V_{\rm cs} =  K_{\rm cs}^{i \ov {j}} \, (D_i W) \, (D_{\ov {j}} \ov{W}) \qquad \text{and}\qquad V_{\rm k} =  K^{{A} \ov {B}} \, (D_{A} W) \, (D_{\ov {B}} \ov{W}) -3 |W|^2\,.
\eea
Moduli stabilization in 4D type IIB effective supergravity models follows a two-step strategy. First, one fixes the complex structure moduli $U^i$ and the axio-dilaton $S$ by the leading order flux superpotential $W_{\rm flux}$ induced by usual S-dual pair of the 3-form fluxes $(F_3, H_3)$. This demands solving the following supersymmetric flatness conditions:
\bea
& & D_i W_{\rm flux} = 0 = D_{\ov {i}} \ov{W}_{\rm flux}, \qquad D_{S} W_{\rm flux} = 0 = D_{\ov {S}} \ov{W}_{\rm flux}.
\label{UStab}
\eea
At this leading order no-scale structrue protects the K\"ahler moduli $T_\alpha$ which subsequently remain flat, and can be stabilized via including other sub-leading contributions to the scalar potential, e.g. those induced via the non-perturbative corrections in the holomorphic superpotential $W$ or the other (non-)perturbative corrections arising from the whole series of $\alpha^\prime$ and string-loop ($g_s$) corrections.

The LVS scheme of moduli stabilization considers a combination of perturbative $(\alpha^\prime)^3$ corrections to the K\"ahler potential ($K$) and a non-perturbative contribution to the superpotential $(W)$ which can be generated by using rigid divisors, such as shrinkable del-Pezzo 4-cycles, or by rigidifying non-rigid divisors using magnetic fluxes \cite{Bianchi:2011qh, Bianchi:2012pn, Louis:2012nb}. The minimal LVS construction includes two K\"ahler moduli corresponding to a so-called Swiss-Cheese like volume form of the CY threefold given as\footnote{Ref. \cite{AbdusSalam:2020ywo} has shown that LVS moduli fixing can be realized also for generic cases where the CY threefold does not have a Swiss-cheese structure.}:
\bea
{\cal V} = \frac{k_{bbb}}{6} \, (t^b)^3 + \frac{k_{sss}}{6} \, (t^s)^3 \, ,
\eea
where $k_{\alpha\beta\gamma}$ denotes the triple intersection number on the CY threefold, and the 2-cycle volume moduli $t^{\alpha}$ are related to the 4-cycle volume moduli $\tau_\alpha$ via $\tau_\alpha = \partial_{t^\alpha} {\cal V}$. Subsequently one has the following Swiss-Cheese like volume form\footnote{Given that the CY threefold has a Swiss-cheese form, one can always find a basis of divisors such that the only non-vanishing intersection numbers are $k_{bbb}$ and $k_{sss}$, which leads to the relation $t^s = - \sqrt{2\tau_s/k_{sss}}$. Here, the minus sign is dictated from the K\"ahler cone conditions as the so-called `small' divisor $D_s$ in this Swiss-cheese CY is an exceptional 4-cycle.}:
\bea
{\cal V} =  \gamma_b \, \, \tau_{b}^{3/2} - \gamma_s\, \, \tau_{s}^{3/2} \,,
\eea
where $\gamma_b = \frac{1}{3} \sqrt{\frac{2}{k_{bbb}}}$ and $\gamma_s = \frac{1}{3} \sqrt{\frac{2}{k_{sss}}}$. The K\"ahler potential including $\alpha'^3$ corrections takes the form \cite{Becker:2002nn}:
\bea
\label{eq:K}
K = -\ln\left[-i\int \Omega\wedge\bar{\Omega}\right]-\ln\left[-\,i\,(S-\bar{S})\right]-2\ln{\cal Y}, \nonumber
\eea
where 
\bea
{\cal Y} = {\cal V}\,+\frac{\xi}{2}\left(\frac{S-\bar{S}}{2i}\right)^{3/2},
\eea
and $\Omega$ denotes the nowhere vanishing holomorphic 3-form which depends on the complex-structure moduli, while the CY volume ${\cal V}$ receives a shift through the $\alpha'^3$ corrections encoded in the parameter $\xi=-\frac{\chi(X)\,\zeta(3)}{2\,(2\pi)^3}$ where $\chi(X)$ is the CY Euler characteristic and $\zeta(3)\simeq 1.202$.

Further, the presence of a `diagonal' del-Pezzo divisor corresponding to the so-called `small' $4$-cycle of the CY threefold induces the superpotential with a non-perturbative effect of the following form:
\bea
W= W_0 + A_s\, e^{- i\, a_s\, T_s}\,,
\label{eq:Wnp-n}
\eea
where after fixing $S$ and the $U$-moduli, the flux superpotential can effectively be considered as constant: $W_0=\langle W_{\rm flux}\rangle$. In addition, the pre-factor $A_s$ can generically depend on the complex-structure moduli which after the first-step of the supersymmetric moduli stabilization can be considered as a parameter. Moreover, without any loss of generality, we consider $W_0$ and $A_s$ to be a real quantity. Subsequently the leading order pieces in the large volume expansion are collected in three types of terms \cite{Balasubramanian:2005zx}:
\bea
\label{eq:VlvsSimpl}
& & V \simeq \frac{{\cal C}_{\alpha'}}{{\cal V}^3} + {\cal C}_{\rm np1}\,\frac{\tau_s}{{\cal V}^2}\, e^{- a_s \tau_s} \cos\left(a_s \,c_s\right) + {\cal C}_{\rm np2}\,\frac{ \sqrt{\tau_s}}{{\cal V}}\, e^{-2 a_s \tau_s},
\eea
where
\bea
& & \hskip-1.5cm {\cal C}_{\alpha'} = \frac{3 \kappa\,\hat\xi |W_0|^2}{4}\,, \quad {\cal C}_{\rm np1} = 4 \kappa\, a_s |W_0| |A_s|\,, \quad {\cal C}_{\rm np2} = 4 \kappa\, a_s^2 |A_s|^2 \sqrt{2 k_{sss}}\,, \quad \kappa = \frac{g_s}{8\pi}\,e^{K_{\rm cs}}. 
\eea
This potential (\ref{eq:VlvsSimpl}) results in exponentially large $\langle{\cal V}\rangle$ determined by
\bea
\label{eq:}
& & \hskip-0.75cm \langle {\cal V} \rangle \simeq \frac{{\cal C}_{\rm np1} \sqrt{\langle \tau_s \rangle}}{2\,{\cal C}_{\rm np2}}\, e^{a_s \langle \tau_s \rangle}, \qquad \qquad \langle \tau_s \rangle \simeq \hat\xi^{2/3} \, \left(\frac{9\,k_{sss}}{8}\right)^{1/3}.
\eea
Here $k_{sss} = 9 - n$ is the degree of the dP$_n$ divisor such that dP$_0 = {\mathbb P}^2$. Thus, after fixing ${\cal V}$ and $\tau_s$, the LVS models based on Swiss-cheese CY with $h^{1,1}({\rm CY}) \geq 3$ have leading order flat directions which can act as promising inflaton candidate with a sub-leading potential.

Given that non-perturbative contributions to the superpotential are not generically present in a given construction, an alternative to the standard LVS has been recently proposed. It has been found that the inclusion of the so-called {\it log-loop} effects \cite{Antoniadis:2018hqy,Antoniadis:2018ngr,Antoniadis:2019doc,Antoniadis:2019rkh,Antoniadis:2020ryh,Antoniadis:2020stf} along with the BBHL corrections \cite{Becker:2002nn}, can result in an exponentially large VEV for the overall volume of the CY threefold just by using the perturbative effects only. With the inclusion of such `log-loop' corrections, the K\"ahler potential $K$ is modified through ${\cal Y}$ which is given as ${\cal Y} = {\cal V} +  \frac{\hat\xi}{2} + \hat\eta\left(\ln{\cal V} -1\right)$, where $\hat\xi = \xi/g_s^{3/2}$, and $\hat\eta = -g_s^2\hat\xi\,\zeta[2]/\zeta[3]$ \cite{Leontaris:2022rzj, Bera:2024zsk,Bera:2024ihl,Leontaris:2025hly}. Now, considering the tree-level superpotential $W_0$ for complex structure moduli and axio-dilaton stabilization, one gets the following scalar potential for perturbative LVS,
\bea
\label{eq:pheno-potV2}
& & \hskip-0.5cm V_{\rm pLVS} = \frac{3\, \kappa\, \hat\xi}{4\, {\cal V}^3}\, |W_0|^2 + \frac{3 \, \kappa\, \hat\eta\, (\ln{\cal V} - 2)}{2{\cal V}^3}\,|W_0|^2,
\eea
which results in an exponentially large VEV for the overall volume modulus determined as,
\bea
\label{eq:pert-LVS}
& & \hskip-0.5cm \langle {\cal V} \rangle \simeq e^{a/g_s^2 + b}, \qquad a = \frac{\zeta[3]}{2 \zeta[2]} \simeq 0.37, \qquad b = \frac73,
\eea
which similar to the standard LVS, corresponds to an AdS minimum with large $\langle {\cal V} \rangle$, and hence justifying the name  ``perturbative LVS". In fact, for a range of string coupling values $g_s = \{0.1, 0.2, 0.25\}$, Eq.~(\ref{eq:pert-LVS}) results in $\langle {\cal V} \rangle \simeq \{7.6 \cdot 10^{16}, 95595, 3567\}$ respectively!

\subsection{Three classes of LVS inflationary models}

The minimal LVS scheme of moduli stabilization fixes the CY volume ${\cal V}$ along with a small modulus $\tau_s$ controlling the volume of an exceptional del-Pezzo divisor. Therefore any LVS models with 3 or more K\"ahler moduli, $h^{1,1}\geq 3$, can generically have flat directions at leading order. These flat directions are promising inflaton candidates with a potential generated at sub-leading order. Based on the geometric nature of the inflaton field and the source of inflaton potential, there are three popular inflationary models based on LVS mechanism to fix the overall volume of the internal CY threefold. These are briefly summarized as below:
\begin{itemize}

\item
(Loop) Blow-up inflation  \cite{Conlon:2005jm, Bond:2006nc, Cicoli:2017shd, Bansal:2024uzr}: 

In this inflationary scenario, the inflaton field is the volume of a (diagonal) del-Pezzo divisor wrapped by an ED3-instanton or supporting gaugino condensation. In addition, the CY has to feature at least one additional ddP$_n$ divisor to realize LVS, and the inflationary potential is generated by non-perturbative superpotential contributions.

Thus, the blow-up inflation model is a three field model based on a two-hole Swiss-cheese CY threefold. Such a CY threefold has two diagonal del-Pezzo divisors, say $D_{s_1}$ and $D_{s_2}$, with volumes $\tau_{s_1}$ and $\tau_{s_2}$ along with the `big' divisor volume $\tau_b$ controlling the overall size of the CY threefold. Subsequently, the overall volume can be rewritten in terms of these 4-cycle volume moduli as below,
\bea
{\cal V} = \gamma_b\, \tau_b^{3/2}- \gamma_{s_1} \, \, \tau_{s_1}^{3/2} - \gamma_{s_2}\, \, \tau_{s_2}^{3/2} \, ,
\eea
where $\gamma_\alpha = \frac{1}{3} \sqrt{\frac{2}{k_{\alpha\alpha\alpha}}}$ for $\alpha \in \{b, s_1, s_2\}$, which means that $\gamma_{\rm ddP_n} = \frac{1}{3} \sqrt{\frac{2}{9-n}}$ for a diagonal del-Pezzo (ddP$_n$) divisor. Moreover, the presence of two rigid diagonal del-Pezzo divisors facilitates the following superpotential,
\bea
W= W_0 + \sum_{\alpha = 1}^2 A_{s_\alpha}\, e^{- i\, a_{s_\alpha}\, T_{s_\alpha}}\,,
\label{eq:Wnp-2n}
\eea

With these pieces of information at hand, the scalar potential of  3 K\"ahler moduli takes the following form \cite{Conlon:2005jm, Cicoli:2017shd}:
\bea
V \simeq \frac{{\cal C}_{\alpha'}}{{\cal V}^3} + \sum_{\alpha=1}^2 \left({\cal C}_{{\rm np1},\alpha}\,\frac{\tau_{s_\alpha}}{{\cal V}^2}\, e^{- a_{s_\alpha} \tau_{s_\alpha}}\, \cos(a_{s_\alpha} c_{s_\alpha}) + {\cal C}_{{\rm np2},\alpha}\,\frac{\sqrt{\tau_{s_\alpha}}}{{\cal V}} \, e^{- 2 a_{s_\alpha}\tau_{s_\alpha}}\right), 
\eea
where
\bea
{\cal C}_{\alpha'} = \frac{3 \kappa\, \hat\xi |W_0|^2}{4}, \quad {\cal C}_{{\rm np1},\alpha} = 4 \kappa\, a_{s_\alpha} |W_0| |A_{s_\alpha}|, \quad {\cal C}_{{\rm np2},\alpha} = \frac{8 \kappa}{3\gamma_{s_\alpha}}\, a_{s_\alpha}^2 |A_{s_\alpha}|^2, \quad \kappa = \frac{g_s}{8\pi}\, e^{K_{cs}}. \nn
\eea

It has been found that the scalar potential induced by this superpotential can drive inflation effectively by a single field after two of the three moduli are stabilized at their respective LVS minimum \cite{Conlon:2005jm}. In fact, a three-field inflationary analysis has been also presented in \cite{Blanco-Pillado:2009dmu,Cicoli:2017shd} ensuring that one can indeed have trajectories which effectively corresponds to a single field dynamics, in the sense of generating the number of efoldings.

However, it has been recently argued that string-loop corrections are likely to be present generically in all the blow-up inflation models and it is hard to escape such corrections. In such situation, the exponentially suppressed contributions can be negligibly small against the following loop blow-up inflationary potential \cite{Bansal:2024uzr},
\bea
& & V_{\rm LBI}(\phi) \simeq V_0 - \frac{{\cal C}_{\rm loop}}{\phi^{2/3}},
\eea
where $\phi$ is the canonically normalized blow-up divisor volume given as $\tau_s \propto \phi^{4/3}$ and ${\cal C}_{\rm loop}$ involves some volume suppression factor. Naively one can understand the $\phi$ dependence in $V(\phi)$ as the K\"ahler potential contributions due to the Winding-type loop corrections scales as $K_{g_s} \propto -\frac{1}{t^s} \simeq \frac{1}{\sqrt\tau_s}$ where the two-cycle volume and the four-cycle volumes corresponding to  the exceptional del-Pezzo divisor are related as $t^s \simeq - \sqrt\tau_s$.

Nevertheless, we will present the classification results for blow-up inflation in the same spirit of the conventional model proposed in \cite{Conlon:2005jm, Cicoli:2017shd}, and we will also witness in the concrete global model that string-loop corrections are hard to escape and a loop blow-up potential is relatively natural and easier to realize !

\item
Fibre inflation \cite{ ,Cicoli:2016xae, Cicoli:2017axo}

These are $3$-field models based on K3-fibred CY threefolds with at least one diagonal del-Pezzo divisor to support LVS. In this class of models, the typical volume form is given as:
\bea
& & {\cal V} =  \gamma_b \, \, \tau_{b} \, \sqrt{\tau_f} - \gamma_s \, \, \tau_{s}^{3/2}\,, 
\eea
The overall volume ${\cal V}$ and small divisor volume $\tau_s$ are stabilized via the LVS prescription while the volume modulus corresponding to the K3-fibre $\tau_f$ drives the inflation through a sub-leading scalar potential effect encoded as 1-loop correction in the K\"ahler potential. In the Einstein frame, it takes the following form \cite{Berg:2004ek, Berg:2005ja, Berg:2005yu, Berg:2007wt, Cicoli:2007xp},
\bea
\label{eq:KgsE}
& & \hskip-1cm K_{g_s}^{\rm KK} = g_s \sum_\alpha \frac{C_\alpha^{\rm KK} \, t^\alpha_\perp}{\cal V} \,, \qquad K_{g_s}^{\rm W} =  \sum_\alpha \frac{C_\alpha^W}{{\cal V}\, t^\alpha_\cap}\,.
\eea
where $C_\alpha^{\rm KK}$ and $C_\alpha^{\rm W}$ are some complex structure moduli dependent functions which can generically also depend on the open-string moduli. The two-cycle volume moduli $t^\alpha_{\perp}$ denote the transverse distance among the various stacks of the non-intersecting $D7$-brane and $O7$-planes, whilst  $t^\alpha_{\cap}$ denotes the volume of the curve sitting at the intersection loci of the various non-trivially intersecting stacks of $D7$-branes such that the intersecting curve is non-contractible. Let us also mention that it has been argued recently in \cite{Gao:2022uop} that $t^\alpha_\perp$ and $t^\alpha_\cap$ can be generically expressed by some more general homogeneous functions of two-cycle volumes of degree-1. However, we will consider the linear choice of such functions in our current analysis, following the prescription of \cite{Berg:2004ek, Berg:2005ja, Berg:2005yu, Berg:2007wt}.

These KK-type and Winding-type effects induce correction to the scalar potential which take the following form,
\bea
& & V_{g_s}^{\rm KK} = \kappa \, g_s^2 \, \frac{|W_0|^2}{4\,{\cal V}^4} \sum_{\alpha,\beta} {\cal C}_\alpha^{\rm KK} {\cal C}_\beta^{\rm KK} \left(2\,t^\alpha t^\beta - 4\, {\cal V} \,k^{\alpha\beta}\right), \\
& & V_{g_s}^{\rm W} = -2 \kappa \frac{|W_0|^2}{{\cal V}^3} \, \sum_\alpha \frac{{\cal C}_{w_\alpha}}{t^\alpha_\cap}, \nonumber
\eea
Such effects can generically appear at ${\cal O}({\cal V}^{-10/3})$ in the large volume expansion. 

In fact there can be additional loop corrections motivated by the field theoretic computations \cite{vonGersdorff:2005bf,Gao:2022uop} as well as the so-called `log-loop' corrections studied in \cite{Antoniadis:2018hqy,Antoniadis:2018ngr,Antoniadis:2019doc,Antoniadis:2019rkh,Antoniadis:2020ryh,Antoniadis:2020stf}, however we do not include those corrections in the current analysis.

\item
Poly-instanton inflation \cite{Cicoli:2011ct,Blumenhagen:2012ue,Blumenhagen:2012kz}

In the minimalistic formulation, poly-instanton inflation is also a $3$-field LVS model \cite{Cicoli:2011ct, Blumenhagen:2012ue}. This involves exponentially suppressed corrections appearing on top of the usual non-perturbative superpotential effects arising from the E3-instantons or gaugino condensation wrapping suitable rigid cycles in the CY threefold. In this three-field model, two K\"ahler moduli correspond to volumes of the `big' and `small' 4-cycle (namely $D_b, D_s$) of a Swiss-cheese CY threefold, while the third modulus corresponds to the volume of the so-called `Wilson' divisor $D_W$. These are some $4$-cycles which can be realized as ${\mathbb P}^1$ fibrations over ${\mathbb T}^2$ base \cite{Blumenhagen:2012kz} and it turns out that such a divisor has the corresponding Hodge numbers given as: 
\bea
& & h^{0,0}(D_W)=1, \qquad h^{1,0} (D_W)= 1, \qquad h^{2,0}(D_W) = 0,  
\eea
and $h^{1,0}_+(D_W) = h^{1,0} (D_W)= 1$ for some specific choice of involution. For this model one has a rather peculiar structure in the volume form given as \cite{Blumenhagen:2012kz}:
\bea
& & {\cal V} =  \gamma_b \, \, \tau_{b}^{3/2} - \gamma_s \, \, \tau_{s}^{3/2} - \gamma_s\, \, (\tau_s + \tau_w)^{3/2}\,, 
\eea
where there are following relations among the $2$-cycle and $4$-cycle volume moduli,
\bea
& & \hskip-1cm \tau_b = \frac{1}{2}\, k_{bbb}\, (t^b)^2, \quad \tau_s = \frac{1}{2}\, k_{sss} \, (t^s - t^w)^2, \quad \tau_w = - \, \frac{1}{2}\, k_{sss} \, (t^s -\, 2\, t^w)\,.
\eea
Moreover, the inflationary potential is induced by the following superpotential \cite{Blumenhagen:2012ue,Blumenhagen:2012kz},
\be
W= W_0 + A_{s}\, e^{- i\, a_{s}\, \left(T_{s}+ A_w e^{-i\, a_w T_w}\right)} \,, 
\label{eq:Wnp-pn}
\ee
where $a_s = 2\pi/n$ for some $n \in {\mathbb N}$ while $a_w = 2 \pi$. Further, the quantities $W_0, A_s$ and $A_w$ generically depend on the complex-structure moduli, and can be considered as parameter after the first step of (complex-structure) moduli stabilization.

\end{itemize}


\section{Classifying divisor topologies for LVS inflation}
\label{sec_topo-classification}
We consider the CY threefolds arising from the four-dimensional reflexive polytopes listed in the Kreuzer-Skarke (KS) database \cite{Kreuzer:2000xy}, and classify the divisors according to their relevance for the building of phenomenological models with the aim of explicit orientifold constructions. In this regard, a collection of the various topological data of CY threefolds with $1\le h^{1,1}({\rm CY}) \le 6 $ is available in the Altman-Gray-He-Jejjala-Nelson (AGHJN) database of \cite{Altman:2014bfa,Altman:2021pyc} that can be directly used for performing model building related studies. For our analysis, in particular, we will mainly use the following input data from AGHJN database:
\begin{itemize}
\item GLSM charges
\item Stanley-Reisner (SR) ideal
\item Fundamental group
\item Intersection tensor $\kappa_{ijk}$
\item Second Chern-class $c_2({\rm CY})$ of the CY threefold
\end{itemize}

\subsection{Methodology}
Let us start by briefly reviewing the generic methodology for analyzing the divisor topologies which is widely adopted for scanning useful CY geometries suitable for phenomenology, e.g. see \cite{Cicoli:2021dhg,Shukla:2022dhz,Cicoli:2023njy}. Subsequently, we will continue following the same in our current analysis. For a given CY geometry, our main focus is limited to:
\begin{itemize}
\item 
Looking at the topology of the so-called `coordinate divisors' $D_i$ which are defined through setting the toric coordinates to zero, i.e. $x_i = 0$. This means that there is a possibility of missing a huge number of divisors, e.g. those which could arise via considering some linear combinations of the coordinate divisors, and some of such may have interesting properties. However, it is hard to perform an exhaustive analysis that includes all the effective divisors of a given CY threefold.

\item
Focusing on scans using `favorable' triangulations (Triang) and `favorable' geometries (Geom) for a given polytope. This could be justified in the sense that for non-favorable CY threefolds, the number of toric divisors in the basis is less than $h^{1,1}(X)$, and subsequently there is always at least one coordinate divisor which is non-smooth, and one usually excludes such spaces from the scan. However, the number of such CY geometries is almost negligible. 

\item 
For the purpose of scanning the divisor topologies, our central task is to consider favorable CY geometries with $1\le h^{1,1}({\rm CY}) \le 6 $. This results in the need to calculate the Hodge diamonds of around a million divisors as we present in Table \ref{tab_PTG}.  

\item 
We also note that in our current analysis we will consider the CY threefolds which have trivial fundamental groups only, which results in discarding 14 favorable CY geometries.

\end{itemize} 

\noindent

\begin{table}[H]
  \centering
 \begin{tabular}{||c||c|c|c||c|c|c||c|c||}
\hline
 $h^{1,1}$ & Polytope  & Triang. & Geom. & Fav.  & Fav. & Fav. & Fav. & Number of \\
&  & & & Poly. & Triang. & Geom. & Geom$^\ast$  & Divisors \\
 \hline
 1 & 5 & 5 & 5 & 5 & 5 & 5 & 4 & 20 \\
 2 & 36 & 48 & 39 & 36 & 48 & 39 & 37 & 222 \\
 3 & 244 & 526 & 306 & 243 & 525 & 305 & 300 & 2100 \\
 4 & 1197 & 5398 & 2014 & 1185 & 5330 & 2000 & 1994 & 15952\\
 5 & 4990 & 57132 & 13635 & 4897 & 56714 & 13494 & 13494 & 121446 \\
 6 & 17101 & 589025 & 85679 & 16608 & 584281 & 84525 & 84525 & 845250 \\
 \hline
 Tot. & 23573 & 652134 & 101678 & 22974 & 646903 & 100368 & 100354 & 984990 \\
 \hline
  \end{tabular}
\caption{Number of triangulations and CY geometries arising from the KS database \cite{Kreuzer:2000xy}.}
\label{tab_PTG}
\end{table}

\noindent
Table \ref{tab_PTG} presents the number of (favorable) polytopes along with the corresponding (favorable) triangulations and (favorable) geometries for a given $h^{1,1}({\rm CY})$ in the range $1 \le h^{1,1}({\rm CY}) \le 6$. The numbers listed in Table \ref{tab_PTG} corresponds to the latest results in \cite{Altman:2021pyc} which may have a slight difference from the initial estimates presented in \cite{Altman:2014bfa,Altman:2017vzk}. In addition, Fav-Geom$^\ast$ denotes those favorable CY geometries that have a trivial fundamental group as motivated in \cite{Shukla:2022dhz}, and the last column presents the total number of divisor topologies corresponding to a given $h^{1,1}({\rm CY})$.

For a generic divisor ($D$) of the CY threefold $X$, there are only four independent Hodge numbers in the Hodge diamond, which are denoted as $h^{0,0}, h^{1,0}, h^{2,0}$ and $h^{1,1}$. Two of these can be computed from the Euler characteristics $\chi(D)$ and the Arithmetic genus $\chi_{_h}(D)$ of the divisor $D$ which can be computed via knowing the second Chern class of the $X$ along with the classical triple intersection numbers $\kappa_{ijk}$. The explicit formulae connecting these quantities can be given as below (e.g. see \cite{Blumenhagen:2008zz,Collinucci:2008sq, Bobkov:2010rf,Cicoli:2016xae}),
\bea
\label{eq:chi-chih}
& & \hskip-1.05cm \chi(D) = 2 h^{0,0} - 4 h^{1,0} + 2 h^{2,0} + h^{1,1} \\
& & = \int_{X} \hat{D} \wedge \hat{D} \wedge \hat{D} + \int_{X} c_2(X) \wedge \hat{D} = \kappa_0(D) + \Pi(D), \, \nonumber\\
& & \hskip-1.2cm \chi_{_h}(D) = h^{0,0} - h^{1,0} + h^{2,0} \nonumber\\
& & = \frac{1}{6} \int_{X} \hat{D} \wedge \hat{D} \wedge \hat{D} + \frac{1}{12} \int_{X} c_2(X) \wedge \hat{D} = \frac{1}{6}\kappa_0(D) + \frac{1}{12}\Pi(D),
\eea
where the second Chern class of the CY threefolds as $c_2(X)$ and, once again, $\hat{D}$ denotes the 2-forms dual to the divisor class, and we define
\bea
& & \kappa_0(D) = \int_{X} \hat{D} \wedge \hat{D} \wedge \hat{D}, \quad \qquad \Pi(D) = \int_{X}  c_2(X) \wedge \hat{D}.
\eea
Thus, after knowing $\chi(D)$ and $\chi_{_h}(D)$ of a divisor using the second Chern class and the triple intersection numbers, one is practically left with computing only two out of the four Hodge numbers. We follow the prescription in \cite{Shukla:2022dhz} to compute the Hodge numbers of all the toric divisors using \texttt{HodgeDiamond}  and  \texttt{Lambda0CotangentBundle} modules of the \texttt{cohomCalg} package  \cite{Blumenhagen:2010pv,Blumenhagen:2011xn}, in combination with the $\kappa_0(D)$ and $\Pi(D)$ data computed from the second Chern class and the triple intersection numbers.

Knowing the Hodge numbers of the coordinate divisors determines the following four topological quantities which remain at the core of global model building and can be subsequently utilized at various stages,
\bea
\label{eq:topo-via-hodgenumbers}
& & \chi_{_h}(D) = h^{0,0}(D) - h^{1,0}(D) + h^{2,0}(D), \, \\
& & \chi(D) = 2 h^{0,0}(D) - 4 h^{1,0}(D) + 2 h^{2,0}(D) + h^{1,1}(D),\, \nonumber\\
& & \kappa_0(D) 
= 10 h^{0,0}(D) - 8 h^{1,0}(D) + 10 h^{2,0}(D) - h^{1,1}(D),\, \nonumber\\
& & \Pi(D) 
= -8 h^{0,0}(D) + 4 h^{1,0}(D) - 8 h^{2,0}(D) + 2 h^{1,1}(D).\, \nonumber
\eea
Using these four relations, one can invoke various special cases, some of which could corresponds to interesting topologies. For example, one can consider the following four cases:
\begin{itemize}
\item
$D_{\chi_{h}}$: Divisors with vanishing Arithmetic genus $\chi_{_h}$.

\item
$D_\chi$: Divisors with vanishing Euler characteristics $\chi$.

\item
$D_\kappa$: Divisors with vanishing self cubic-intersections $\kappa_0$.

\item
$D_\Pi$: Divisors with vanishing $\Pi$.

\end{itemize}
\noindent
We find that there are 8 divisors topologies of $D_{\chi_{h}}$ type, 4 divisors topologies of $D_\chi$ type, 7 divisors topologies of $D_\kappa$ type, and only 2 divisors topologies of $D_\Pi$ type. In fact, the divisors of vanishing $\Pi = 0$ are those corresponding to a del-Pezzo surface of degree 6, i.e. a dP$_3$ and an exact Wilson divisor invoked for poly-instanton correction in \cite{Blumenhagen:2012kz}.

\begin{table}[H]
\centering
\begin{tabular}{|c||c||c|c|c|c|c|c||c|} 
\hline
Sr. No. & Divisor Topologies: & $f_1$ & $f_2$ & $f_3$ & $f_4$ & $f_5$ & $f_6$ & $f_{\rm all}$ \\
 & $\{h^{0,0}, h^{1,0}, h^{2,0}, h^{1,1}\}$ &&&&&& & \\
\hline
\hline
1 & Rigid: \{1,0,0,$n$\}  & 0 & 33 & 617 & 7339 & 68850 & 549409 & 626248 \\
2 & Wilson: \{1,$m$,0,$n$\} & 0 & 5 & 91 & 754 & 6079 & 41255 & 48184 \\
3 & Non-rigid: \{1,0,$m$,$n$\} & 20 & 184 & 1392 & 7859 & 46517 & 254586 & 310558\\
\hline
Total &  & 20 & 222 & 2100 & 15952 & 121446 & 845250 & 984990 \\
\hline
\end{tabular}
\caption{Collection of three classes of topologies corresponding to the so-called Rigid, Wilson and Non-rigid divisors. Here $f_p$ denotes the respective frequencies of a particular divisor topology for a given $p = h^{1,1}(\rm CY)$ while $m$ and $n$ are some non-zero integers.}
\label{tab_dPn-topo-list}
\end{table}

\begin{table}[h]
\centering
\begin{tabular}{|c||c||c|c|c|c||c|c|c|c|c|c||c|} 
\hline
Sr. & Divisor & $f_1$ & $f_2$ & $f_3$ & $f_4$ & $f_5$ & $f_6$ & $f_{\rm all}$ \\
No. & Topology &&&&&& & \\
\hline
1 & \{1,0,0,1\} & 0 & 8 & 59 & 372 & 2410 & 14490 & 17339 \\
2 & \{1,0,0,2\} & 0 & 4 & 103 & 999 & 9224 & 69022 & 79352 \\
3 & \{1,0,0,3\} & 0 & 0 & 4 & 160 & 2360 & 25070 & 27594 \\
4 & \{1,0,0,4\} & 0 & 0 & 4 & 152 & 2441 & 26772 & 29369 \\
5 & \{1,0,0,5\} & 0 & 0 & 6 & 103 & 1586 & 17335 & 19030 \\
6 & \{1,0,0,6\} & 0 & 0 & 9 & 198 & 2574 & 25842 & 28623 \\
7 & \{1,0,0,7\} & 0 & 2 & 21 & 250 & 2705 & 24617 & 27595\\
8 & \{1,0,0,8\} & 0 & 4 & 67 & 714 & 5988 & 49058 & 55831 \\
9 & \{1,0,0,9\} & 0 & 5 & 75 & 673 & 5772 & 45610 & 52135 \\
\hline
10 & \{1,0,0,10\} & 0 & 0 & 54 & 927 & 8983 & 66262 & 76226 \\
11 & \{1,0,0,11\} & 0 & 1 & 35 & 516 & 5364 & 44486 & 50402 \\
\hline
12 & \{1,0,1,20\} & 0 & 24 & 322 & 1879 & 10745 & 56702 & 69672 \\
13 & \{1,0,2,30\} & 0 & 30 & 235 & 1145 & 5851 & 29568 & 36829 \\
14 & \{1,1,0,2\} & 0 & 0 & 17 & 214 & 1940 & 13750 & 15921 \\
\hline
15 & \{1,0,1,$m$\} & 0 & 16 & 245 & 1803 & 12877 & 77900 & 92841 \\
16 & \{1,1,0,$n$\} & 0 & 0 & 0 & 19 & 497 & 5535 & 6051 \\
\hline
\hline
Total &  & 0 & 94 & 1256 & 10124 & 81317 & 592019 & 684810 \\
\hline
\end{tabular}
\caption{Collection of 16 frequently appearing divisor topologies classified with their Hodge numbers $\{h^{0,0}, h^{1,0}, h^{2,0}, h^{1,1}\}$. Here $f_p$ denotes the respective frequencies of a particular divisor topology for a given $p = h^{1,1}(\rm CY)$, and $m \neq 20$ and $n \ne 2$ are some non-zero integers. This table shows that approximately 70\% of the total divisor topologies fall in just 16 categories.}
\label{tab_dPn-topo-list}
\end{table}

\subsection{Diagonal del-Pezzo divisors and LVS}
The role of divisor topologies in the LVS context can be appreciated by noting that the Swiss-cheese structure of the CY volume can be correlated with the presence of del-Pezzo (dP$_n$) divisors $D_s$. These dP$_n$ divisors are defined for $ 0 \leq n \leq 8$ having degree $d = 9 -n$ and $h^{1,1} = 1 + n$, such that dP$_0$ is a ${\mathbb P}^2$ and the remaining 8 del-Pezzo's are obtained by blowing up eight generic points inside ${\mathbb P}^2$. Subsequently, the Hodge diamond corresponding to the del-Pezzo surface $\mathrm{dP}_n$ turns out to be the following.

\bea
\label{eq:HodgeDiamond}
{\rm dP}_n &\equiv& \begin{tabular}{ccccc}
    & & 1 & & \\
   & 0 & & 0 & \\
  0 & & $n+1$ & & 0\\
   & 0 & & 0 & \\
    & & 1 & & \\
  \end{tabular}; \qquad {\rm dP}_0 = {\mathbb P}^2 \, \, {\rm and} \,\, 1 \le n \le 8, \nonumber
\eea
where $\chi_h({\rm dP}_n) = 1, \, \chi({\rm dP}_n) = n+3, \, \, \kappa_0({\rm dP}_n) = 9-n$, and $\Pi({\rm dP}_n) = 2n-6$. However, merely looking at the Hodge numbers of a divisor $D_s$ does not fully ensure that it is a del-Pezzo surface. In addition, the following necessary conditions needs to be satisfied \cite{Cicoli:2011it}:
\be
\label{eq:dP}
\int_X D_s^3 = k_{sss} > 0\, , \qquad \int_X D_s^2 \, D_i \leq 0 \qquad \forall \, i \neq s \,.
\ee
Here the self-triple-intersection number $k_{sss}$ corresponds to the degree of the del-Pezzo 4-cycle dP$_n$ where $k_{sss} = 9 - n > 0$, which is always positive as $n \leq 8$ for del Pezzo surfaces. 

\noindent
\begin{table}[h]
\centering
\begin{tabular}{|c||c|c|c|c|c||c|c|c|c|c|c||c|} 
\hline
&&&&&&&&&&&& \\
 dP$_n$ & $h^{p,q}$  & $\chi_{_h}$ & $\chi$ & $\kappa_0$ & $\Pi$ & $f_1$ & $f_2$ & $f_3$ & $f_4$ & $f_5$ & $f_6$  & $f_{\rm all}$ \\
Type&&&&&&&&&&&& \\
\hline
&&&&&&&&&&&& \\
 dP$_0$ & \{1,0,0,1\} & 1 & 3 & 9 & -6 & 0 & 8 & 59 & 372 & 2410 & 14490 & 17339 \\
 dP$_1, {\mathbb F}_0$ & \{1,0,0,2\} & 1 & 4 & 8 & -4 & 0 & 4 & 91 & 878 & 8038 & 59707 & 68718 \\
 dP$_2$ & \{1,0,0,3\} & 1 & 5 & 7 & -2 & 0 & 0 & 4 & 155 & 2242 & 23556 & 25957 \\
 dP$_3$ & \{1,0,0,4\} & 1 & 6 & 6 & 0 & 0 & 0 & 4 & 144 & 2271 & 24658 & 27077 \\
 dP$_4$ & \{1,0,0,5\} & 1 & 7 & 5 & 2 & 0 & 0 & 2 & 55 & 947 & 12079 & 13083 \\
 dP$_5$ & \{1,0,0,6\} & 1 & 8 & 4 & 4 & 0 & 0 & 9 & 184 & 2190 & 21898 & 24281 \\
 dP$_6$ & \{1,0,0,7\} & 1 & 9 & 3 & 6 & 0 & 2 & 21 & 239 & 2459 & 21294 & 24015 \\
 dP$_7$ & \{1,0,0,8\} & 1 & 10 & 2 & 8 & 0 & 4 & 67 & 689 & 5462 & 43098 & 49320 \\
 dP$_8$ & \{1,0,0,9\} & 1 & 11 & 1 & 10 & 0 & 5 & 71 & 597 & 4692 & 35532 & 40897 \\
\hline
\hline
Total &&&&&& 0 & 23 & 328 & 3313 & 30711 & 256312 & 290687 \\
\hline
\end{tabular}
\caption{Del-Pezzo divisor topologies satisfying Eq. (\ref{eq:dP}). Here dP$_0={\mathbb P}^2$ and ${\mathbb F}_0 = {\mathbb P}^1\times{\mathbb P}^1$.}
\label{tab_dPn-topo-list}
\end{table}

\noindent
Furthermore, in order to have the Swiss-Cheese structure, one needs at least one diagonal del-Pezzo divisor in the CY threefold. So, one imposes the so-called `diagonality' condition on such a del-Pezzo divisor $D_s$ via the following relation satisfied by the triple intersection numbers \cite{Cicoli:2011it, Cicoli:2018tcq}:
\bea
\label{eq:diagdP}
k_{sss} \, \, k_{s i j } = k_{ss i} \, \, k_{ss j} \, \qquad \qquad \forall \, \, \, i, j.
\eea
It turns out that whenever this diagonality condition is satisfied, there exists a basis of coordinates divisors such that the volume of each of the 4-cycles $D_s$ becomes a complete-square quantity as illustrated from the following relations:
\bea
\tau_s = \frac{1}{2}\, k_{s ij} t^i \, t^j = \frac{1}{2 \, k_{sss}}\, k_{ssi} \, k_{s s j} t^i \, t^j = \frac{1}{2 \, k_{sss}}\, \left(k_{ss i} \,t^i \, \right)^2\,.
\eea
Subsequently what happens is that one can always shrink such a `diagonal' del-Pezzo ddP$_n$ to a point-like singularity by squeezing it along a single direction. Imposing the diagonality condition for the del-Pezzo analysis presented in Table \ref{tab_dPn-topo-list} leads to the Table \ref{tab_ddPn-topo-list}. In fact, a systematic analysis on counting the CY geometries which could support (standard) LVS models, in the sense of having at least one diagonal del-Pezzo divisor, has been performed for $ 1 \le h^{1,1}({\rm CY}) \leq 5$ in \cite{Cicoli:2021dhg}, where it has been observed that the CY threefolds arising from the KS dataset do not have a `diagonal' dP$_n$ divisor for $ 1 \leq n \leq 5$. Subsequently, it has been conjectured to be true for all the CY geometries arising from the KS database. In the current analysis, we find the conjecture to be true for $h^{1,1} = 6$ as well. The specifics of this result are collected in Table \ref{tab_ddPns-GstarM}. 

\noindent
\begin{table}[h]
\centering
\begin{tabular}{|c||c|c|c|c|c||c|c|c|c|c|c||c|} 
\hline
&&&&&&&&&&&& \\
 ddP$_n$ & $h^{p,q}$  & $\chi_{_h}$ & $\chi$ & $\kappa_0$ & $\Pi$ & $f_1$ & $f_2$ & $f_3$ & $f_4$ & $f_5$ & $f_6$ & $f_{\rm all}$ \\
Type&&&&&&&&&&&& \\
\hline
 dP$_0={\mathbb P}^2$ & \{1,0,0,1\} & 1 & 3 & 9 & -6 & 0 & 8 & 59 & 372 & 2410 & 14490 & 17339 \\
 d${\mathbb F}_0$ & \{1,0,0,2\} & 1 & 4 & 8 & -4 & 0 & 2 & 16 & 144 & 944 & 5444 & 6550 \\
 ddP$_1$ & \{1,0,0,2\} & 1 & 4 & 8 & -4 & 0 & 0 & 0 & 0 & 0 & 0 & 0 \\
 ddP$_2$ & \{1,0,0,3\} & 1 & 5 & 7 & -2 & 0 & 0 & 0 & 0 & 0 & 0 & 0 \\
 ddP$_3$ & \{1,0,0,4\} & 1 & 6 & 6 & 0 & 0 & 0 & 0 & 0 & 0 & 0 & 0 \\
 ddP$_4$ & \{1,0,0,5\} & 1 & 7 & 5 & 2 & 0 & 0 & 0 & 0 & 0 & 0 & 0 \\
 ddP$_5$ & \{1,0,0,6\} & 1 & 8 & 4 & 4 & 0 & 0 & 0 & 0 & 0 & 0 & 0 \\
 ddP$_6$ & \{1,0,0,7\} & 1 & 9 & 3 & 6 & 0 & 2 & 17 & 109 & 624 & 1631 & 2383 \\
 ddP$_7$ & \{1,0,0,8\} & 1 & 10 & 2 & 8 & 0 & 4 & 40 & 277 & 827 & 2999 & 4147 \\
 ddP$_8$ & \{1,0,0,9\} & 1 & 11 & 1 & 10 & 0 & 5 & 39 & 157 & 407 & 1330 & 1938 \\
\hline
\hline
Total &&&&&& 0 & 21 & 171 & 1059 & 5212 & 25894 & 32357 \\
\hline
\end{tabular}
\caption{Diagonal del-Pezzo (ddP$_n$) divisors satisfying the diagonality condition Eq.~(\ref{eq:diagdP}).}
\label{tab_ddPn-topo-list}
\end{table}

\noindent

\begin{table}[H]
	\centering
	\hskip0.11cm \begin{tabular}{|c||c||c|c|c|c|c|c||c|}
		\hline
		$h^{1,1}$ &  Geom$^*$ & $\mathrm{ddP}_0$  & $d{\mathbb F}_0$  & $\mathrm{ddP}_n$ & $\mathrm{ddP}_6$  & $\mathrm{ddP}_7$  & $\mathrm{ddP}_8$  & $n_{\rm LVS}$ \\
		  & ($n_{\rm CY}$) & $({\mathbb P}^2)$ & (${\mathbb P}^1 \times {\mathbb P}^1$) & $1\leq n \leq 5$ &  &  &  & ($\mathrm{ddP}_n\geq 1$) \\
		\hline
		1  & 4 & 0 & 0  & 0  & 0 & 0 & 0 & 0 \\
		2  & 37 & 8 & 2 & 0 & 2  & 4  & 5 & 21 \\
		3  & 300 & 55 & 16 & 0  & 16 & 37 & 34 & 132\\
		4  & 1994 & 304 & 140 & 0 & 97 & 210 & 126 &  750 \\
		5  & 13494 & 2107 & 901  & 0  & 486 & 731 & 374 & 4104 \\
		6  &  84525 & 12709 & 5096  & 0  & 1542 & 2851 & 1232 & 20810 \\
		\hline
        \hline
        Total  & 100354 & 15183 & 6155  & 0  & 2143 & 3833 & 1769 & 25817 \\
        \hline
	\end{tabular}
	\caption{CY geometries with at least one ddP$_n$ divisor (of a given type) as needed for LVS.}
	\label{tab_ddPns-GstarM}
\end{table}

\noindent
Table \ref{tab_ddPns-GstarM} shows that there is a landscape of suitable CY geometries for realizing the standard LVS. In fact, we see that more than 25\% of the CY geometries have at least one diagonal del-Pezzo divisor and hence can be suitable for realizing standard LVS. In this regard, it maybe worth noting that one can still have alternative moduli stabilization schemes realizing an exponentially large CY volume, e.g. using the underlying symmetries of the CY threefold in the presence of a non-diagonal del-Pezzo \cite{AbdusSalam:2020ywo}, and in the framework of the so-called perturbative LVS \cite{Antoniadis:2018hqy,Antoniadis:2019rkh,Leontaris:2022rzj,Leontaris:2023obe,Leontaris:2026mdl,Leontaris:2026sqh, Leontaris:2025hly}.

\subsection{Scanning results for LVS Inflationary models}
For CY threefolds with $h^{1,1}({\rm CY}) \leq 5$, the scanning results for the divisor topologies relevant for various LVS inflation models have been proposed in \cite{Cicoli:2023njy}. We extend this result to $h^{1,1}({\rm CY}) = 6$ which significantly enhances the list of candidate CY threefold.

\subsubsection*{Blow-up inflation}
\label{sec_BI}

The minimal LVS scheme of moduli stabilization fixes the CY volume ${\cal V}$ along with a small modulus $\tau_s$ controlling the volume of an exceptional del-Pezzo divisor. Therefore any LVS model with 3 or more K\"ahler moduli, $h^{1,1}\geq 3$, can generically have flat directions at leading order. These flat directions are promising inflaton candidates with a potential generated at sub-leading order. Blow-up inflation \cite{Conlon:2005jm, Bond:2006nc, Cicoli:2017shd} corresponds to the case where the inflationary potential is generated by non-perturbative superpotential contributions. In this inflationary scenario the inflaton is a (diagonal) del-Pezzo divisor wrapped by an ED3-instanton or supporting gaugino condensation. In addition, the CY has to feature at least one additional ddP$_n$ divisor to realize LVS. 

On these lines, we present the scanning results in Table \ref{tab_blowup-Gstar} corresponding to the number of CY geometries $n_{\rm CY}$ with their suitability for realizing LVS along with resulting in the standard blow-up inflationary potential, in the sense of having at least two ddP divisors, one needed to support LVS and the other one to drive inflation. Table \ref{tab_blowup-Gstar} generalizes the previous result \cite{Cicoli:2023njy} by including the analysis for $h^{1,1}({\rm CY}) = 6$.

\begin{table}[H]
\centering
\hskip0.11cm \begin{tabular}{||c|c||c|c|c|c||c|c||}
\hline
 $h^{1,1}$ &  Geom$^*$ & $n_{\rm ddP}=1$  & $n_{\rm ddP}=2$ & $n_{\rm ddP}=3$  & $n_{\rm ddP}\ge 4$ & $n_{\rm LVS}$ & Blow-up \\
&  $(n_{\rm CY})$ &  &   &  &  &  & inflation \\
 \hline
 1 & 4 & 0 & 0  & 0  & 0 & 0 & 0 \\
 2 & 37 & 21 & 0  & 0  & 0 & 21 & 0 \\
 3 & 300 & 93 & 39  & 0 & 0 & 132 & 39 \\
 4 & 1994 & 465 & 261 & 24 & 0 & 750 & 285 \\
 5 & 13494 & 3128  & 857  & 106 & 13 & 4104 & 976 \\
 6 &  84525  & 16177  & 4218  & 380 & 35 & 20810 & 4633 \\
 \hline
 \hline
Total & 100354  & 19884  & 5375  & 510 & 48 & 25817 & 5933 \\
\hline
  \end{tabular}
\caption{Number of candidate CY geometries for Blow-up inflation in LVS.}
   \label{tab_blowup-Gstar}
 \end{table}

\subsubsection*{Fibre inflation}
\label{sec_FI}

The minimal version of Fibre inflation \cite{Cicoli:2008gp,Cicoli:2016chb} also involves three K\"ahler moduli; two of them are stabilized via the standard LVS procedure and the remaining one can serve as an inflaton candidate in the presence of perturbative corrections to the K\"ahler potential. However, fiber inflation requires a different geometry from the one of blow-up inflation since one needs CY threefolds which are K3 fibrations over a $\mathbb{P}^1$ base. The simplest model requires the addition of a blow-up mode such that the volume can be expressed as:
\begin{equation}
\mathcal{V}=\frac{1}{6}\left(k_{111}(t^1)^3+3k_{233}t^2(t^3)^2\right)=\lambda_b \sqrt{\tau_2}\tau_3- \lambda_s \tau_1^{3/2}.
\end{equation}
The requirement of having a K3 fibred CY threefold with at least a ddP$_n$ divisor for LVS moduli stabilisation is quite restrictive. The corresponding scanning results for the number of CY geometries suitable for realizing fibre inflation are presented in Tab.~\ref{tab_fibre-Gstar} which generalizes the earlier results presented in \cite{Cicoli:2023njy}.

\begin{table}[H]
 \centering
\hskip0.11cm \begin{tabular}{||c|c||c|c||c||c|c||}
\hline
 $h^{1,1}$ &  Geom$^*$ & $n_{\rm LVS}$ & K3 fib.  & $n_{\rm LVS}$ \& K3 fib. &  F.I. & F.I \\
   &  $(n_{\rm CY})$  & &  CY & (Fibre Inflation)  &  \& $D_\Pi$ &  \& $D_{W_\Pi}$ \\
 \hline
 1  & 4 & 0 & 0  & 0  &  0  & 0 \\
 2  & 37 & 21 & 10  & 0  &  0  & 0 \\
 3  & 300 & 132 & 136  & 43 &  0  & 0 \\
 4  & 1994 & 750 & 865 & 171  & 28 & 23 \\
 5  & 13494 & 4104  & 5970  &  951 & 179 & 135 \\
 6  &  84525  & 20810  & 35648  & 3800  & 1074 & 584\\
 \hline
 \hline 
Total  &  100354  & 25817  & 42629  & 4965  & 1281 & 742\\
 \hline
  \end{tabular}
  \caption{Number of candidate CY geometries for fibre inflation in LVS.}
    \label{tab_fibre-Gstar}
 \end{table}

It is worth mentioning that the scanning results presented in Table \ref{tab_fibre-Gstar} are consistent with the previous scans performed in \cite{Cicoli:2016xae, Cicoli:2011it, Cicoli:2023njy}. To be more specific, the number of distinct K3 fibred CY geometries supporting LVS was claimed to be 43 for $h^{1,1}=3$ in \cite{Cicoli:2016xae}, while ref. \cite{Cicoli:2011it} claimed that the number of polytopes giving K3 fibred CY threefolds with $h^{1,1}=4$ and at least one diagonal del Pezzo ddP$_n$ divisor is 158. We also note that Fibre inflation has been recently realized in models based on non Swiss-Cheese CY threefolds as well \cite{Bera:2024ihl,Leontaris:2025hly}.

\subsubsection*{Poly-instanton inflation}

In principle, one should be able to fit the requirements for poly-instanton inflation on top of having LVS in a setup with three K\"ahler moduli. Indeed we find that there are four CY threefold geometries with $h^{1,1}(X) = 3$ in the KS database which have exactly one Wilson divisors and a ${\mathbb P}^2$ divisor. However, as mentioned in \cite{Blumenhagen:2012kz}, in order to avoid all vector-like zero modes to have poly-instanton effects, one should ensure that the rigid divisors wrapped by the ED3-instantons, should have some orientifold-odd $(1,1)$-cycles which are trivial in the CY threefold. Given that ${\mathbb P}^2$ has a single $(1,1)$-cycle, it would certainly not have such additional two-cycles which could be orientifold-odd and then trivial in the CY threefold. Hence one has to look for CY examples with $h^{1,1}(X) \ge 4$ for a viable model of poly-instanton inflation as presented in \cite{Blumenhagen:2012kz,Blumenhagen:2012ue}. In this regard, we present the classification of all the candidate CY geometries relevant for LVS poly-instanton inflation in Table \ref{tab_GwilsonLVS} which generalizes the result presented in \cite{Gao:2013pra,Lust:2013kt,Cicoli:2023njy}. 

\begin{table}[H]
  \centering
 \begin{tabular}{||c|c||c|c|c||c|c|c|c||}
\hline
$h^{1,1}$  & Geom$^*$  & Single & Two & $\ge$ Three  & $n_{\rm LVS}$ & $n_{\rm LVS}$ \& $D_W$ & $n_{\rm LVS}$ \&  $D_{W_\Pi}$   \\
&  $(n_{\rm CY})$ & $D_W$ & $D_W$ & $D_W$  &  & (poly-inst.) & (topol. tamed)     \\
 \hline
 1  & 4 & 0 & 0  & 0 & 0 & 0  & 0    \\
 2  & 37 & 0 & 0 & 0 & 21 & 0  & 0  \\
 3  & 300 & 19 & 0 & 0 & 132 & 4 & 4   \\
 4  & 1994 & 221 & 8 & 0 & 750 & 75  & 63   \\
 5  & 13494 & 1874  & 217  & 43 & 4104 &  660 & 522  \\
 6  & 84525 & 13698  & 2126  & 443 & 20810 & 4011  & 3119  \\
 \hline
 \hline
 Total  & 100354 & 15812  & 2351  & 486 & 25817 & 4750  & 3708  \\
 \hline
  \end{tabular}
\caption{Number of candidate CY geometries for poly-instanton inflation in LVS.}
\label{tab_GwilsonLVS}
\end{table}

\noindent
As a side remark, let us recall that for having poly-instanton corrections to the superpotential one needs to find a Wilson divisor $D_W$ with $h^{2,0}(D_W) = 0$ and $h^{0,0}(D_W)= h^{1,0} (D_W)=  h^{1,0}_+(D_W)=1$ for some specific choice of involution, without any restriction on $h^{1,1}(D_W)$ \cite{Blumenhagen:2012kz}. However, a particular type of Wilson divisor with $h^{1,1}(D_W) = 2$ has been found to be useful regarding control over the higher derivative F$^4$ corrections \cite{Cicoli:2023njy}. In this regard, Tables \ref{tab_GwilsonLVS} and \ref{tab_Wilsonmismatch-Gstar} show the existence of several Wilson divisors which fail to have vanishing $\Pi$ since they have $h^{1,1}(W) \neq 2$.

\noindent
 \begin{table}[H]
  \centering
 \begin{tabular}{||c|c||c|c|c|c||c|c|c|c||}
\hline
 $h^{1,1}$  & Geom$^*$  & At least  & single & two & $\ge$ three  & At least  & single & two & $\ge$ three      \\
&  & one $D_W$ & $D_W$ & $D_W$  & $D_W$  & one $D_{W_\Pi}$ & $D_{W_\Pi}$ & $D_{W_\Pi}$  & $D_{W_\Pi}$      \\
 \hline
 1  & 4 & 0 & 0  & 0 & 0 & 0  & 0 & 0 & 0  \\
 2  & 37 &  0 & 0  & 0 & 0 & 0  & 0 & 0 & 0  \\
 3  & 300 & 19 &  19 & 0  & 0  & 19 & 19  & 0  & 0    \\
 4  & 1994 & 229 & 221  & 8 & 0 & 210 & 202 & 8  & 0    \\
 5  & 13494 & 2134  & 1874  & 217 & 43 &  1764 & 1599  &  154 & 11 \\
 6  & 84525 & 16267  & 13698  & 2126 & 443 &  12249 & 10846  & 1311  &  92\\
 \hline
  \end{tabular}
\caption{CY geometries with divisors $D_W$ and $D_{W_\Pi}$ without demanding a diagonal dP divisor.}
\label{tab_Wilsonmismatch-Gstar}
\end{table}


\section{Finding the candidate CY geometries}
\label{sec_TwoEx-UnifiedLVS}
In this section we present the scanning results of CY geometries which are K3-fibred and have two diagonal del-Pezzo divisor along with a suitable Wilson divisor. After scanning nearly a million of toric divisors corresponding to around 100,000 CY geometries with $1 \leq h^{1,1}({\rm CY}) \leq 6$, we find that there are 61 candidate CY geometries which may be suitable for unifying the LVS inflationary models. The scanning results are presented in Table \ref{tab_unified-LVS}.



\begin{table}[H]
\centering
\hskip0.11cm \begin{tabular}{||c|c||c|c|c|c||c|c||}
\hline
 $h^{1,1}$ &  Geom$^*$ & $n_{\rm LVS}$  & K3-fibred & $D_W$  & $D_{W_\Pi}$ & Unified LVS & Unified LVS \\
&  $(n_{\rm CY})$ &  &  &  &  & (with $D_W$) & (with $D_{W_\Pi}$) \\
 \hline
 1 & 4 & 0 & 0  & 0  & 0 & 0 & 0 \\
 2 & 37 & 21 & 10  & 0  & 0 & 0 & 0 \\
 3 & 300 & 132 & 136  & 19 & 19 & 0 & 0 \\
 4 & 1994 & 750 & 865 & 229 & 210 & {\bf 2} & {\bf 0} \\
 5 & 13494 & 4104  & 5970  & 2134 & 1764 & {\bf 14} & {\bf 6} \\
 6 &  84525  & 20810  & 35648  & 16267 & 12249 & {\bf 45} & {\bf 40} \\
 \hline
 \hline
Total &  100354  & 25817  & 42629  & 18649 & 14242 & 61 & 46 \\
 \hline
  \end{tabular}
\caption{Number of candidate CY geometries for unified (BI+FI+PI) inflation in LVS.}
   \label{tab_unified-LVS}
 \end{table}

\noindent
After scanning the Kreuzer-Skarke CY database \cite{Kreuzer:2000xy} for $ 1 \leq h^{1,1}({\rm CY}) \leq 6$, we have found only two candidate CY geometries with $h^{1,1}({\rm CY})= 4$ and having the preliminary ingredients to realize the desired unified LVS inflation. These geometries correspond to the polytope Ids: 314 and 684 in the AGHJN CY database of \cite{Altman:2014bfa}. However, these two CY geometries have Wilson divisor with non-vanishing $\Pi$'s and hence the corresponding F$^4$-correction is present.

For $h^{1,1}({\rm CY}) = 5$, we have found 14 candidate CY geometries with the preliminary ingredients desired to achieve unified LVS inflation, and among these, there are 6 CY geometries which have Wilson divisor of vanishing $\Pi$ as well. These geometries correspond to the polytope Ids: 2579, 2580, 2583, 3352, 4429, 4686, 4687, 4689, 5087, 5789, 6427 and 6430 in the AGHJN CY database of \cite{Altman:2014bfa}, and we also note that there are multiple CY geometries for polytope Ids 2580 and 4687. We present the toric data for such 14 CY geometries in the Appendix \ref{sec_appendix}. Continuing the scan for $h^{1,1}({\rm CY}) = 6$, we have found 45 candidate CY geometries with the preliminary ingredients desired to achieve unified LVS inflation. 

Now we present a couple of explicit CY orientifold constructions to realize the unified LVS inflationary model.

\subsection{Example 1}
The CY threefold corresponding to the polytope Id: 314 is defined by the following toric data:
\begin{center}
\begin{tabular}{|c|cccccccc|}
\hline
\cellcolor[gray]{0.9}Hyp &\cellcolor[gray]{0.9} $x_1$  &\cellcolor[gray]{0.9} $x_2$  &\cellcolor[gray]{0.9} $x_3$  &\cellcolor[gray]{0.9} $x_4$  &\cellcolor[gray]{0.9} $x_5$ & \cellcolor[gray]{0.9}$x_6$  &\cellcolor[gray]{0.9} $x_7$ &\cellcolor[gray]{0.9} $x_8$      \\
\hline
\cellcolor[gray]{0.9} 4 & 0 & 0 & 1 & 0 & 1 & 0 & 1 & 1  \\
\cellcolor[gray]{0.9} 4 & 0 & 0 & 1 & 1 & 1 & 1 & 0 & 0 \\
\cellcolor[gray]{0.9} 8 & 1 & 0 & 2 & 1 & 2 & 0 & 0 & 2 \\
\cellcolor[gray]{0.9} 8 & 1 & 1 & 2 & 2 & 2 & 0  & 0 & 0 \\
\hline
& dP$_7$  & dP$_7$ &  &  K3 &  & $D_W$ &  &   \\
\hline
\end{tabular}
\end{center}
The Hodge numbers are $(h^{2,1}, h^{1,1}) = (52, 4)$, the Euler number is $\chi=-96$ and the Stanley-Reisner ideal is:
\be
{\rm SR} =  \{x_1 x_2, x_1 x_4, x_1 x_8, x_2 x_4, x_2 x_7, x_4 x_6, x_3 x_5 x_6 x_7, x_3 x_5 x_6 x_8, x_3 x_5 x_7 x_8\} \,. \nn
\ee
The analysis of the divisor topologies using {\it cohomCalg} \cite{Blumenhagen:2010pv, Blumenhagen:2011xn} shows that they can be represented by the following Hodge diamonds:
\bea
\label{eq:HodgeDiamond1}
D_1 &\equiv& \begin{tabular}{ccccc}
    & & 1 & & \\
   & 0 & & 0 & \\
  0 & & 8 & & 0\\
   & 0 & & 0 & \\
    & & 1 & & \\
  \end{tabular} \equiv D_2, \qquad \quad \, \,  D_3 \equiv \begin{tabular}{ccccc}
    & & 1 & & \\
   & 0 & & 0 & \\
  3 & & 36 & & 3 \\
   & 0 & & 0 & \\
    & & 1 & & \\
  \end{tabular}, \\
& & \nonumber\\
D_4 &\equiv& \begin{tabular}{ccccc}
    & & 1 & & \\
   & 0 & & 0 & \\
  1 & & 20 & & 1 \\
   & 0 & & 0 & \\
    & & 1 & & \\
  \end{tabular} \equiv K3, \qquad \quad D_5 \equiv \begin{tabular}{ccccc}
    & & 1 & & \\
   & 0 & & 0 & \\
  3 & & 36 & & 3 \\
   & 0 & & 0 & \\
    & & 1 & & \\
  \end{tabular}, \nonumber\\
& & \nonumber\\
D_6 &\equiv& \begin{tabular}{ccccc}
    & & 1 & & \\
   & 1 & & 1 & \\
  0 & & 6 & & 0 \\
   & 1 & & 1 & \\
    & & 1 & & \\
  \end{tabular} \equiv D_W, \qquad \quad \, \, D_7 \equiv \begin{tabular}{ccccc}
    & & 1 & & \\
   & 3 & & 3 & \\
  0 & & 6 & & 0 \\
   & 3 & & 3 & \\
    & & 1 & & \\
  \end{tabular} \equiv D_8. \nonumber
\eea
Let us make an important observation that this CY threefold has an exchange symmetry given by $\{1 \leftrightarrow 2, 7 \leftrightarrow 8\}$, which is present at the level of SR ideal and the Hodge Diamonds of the corresponding toric divisors. The intersection curves lying at the two toric coordinate divisors are presented in Table \ref{Tab3} where ${\cal C}_g$ denotes a curve of genus $g$ which is given by the Hodge number $\{h^{0,0} = 1, h^{1,0} = g \}$, i.e. ${\cal C}_0 = {\mathbb P}^1, {\cal C}_1 = {\mathbb T}^2$, and $n {\mathbb P}^1 = {\mathbb P}^1 \sqcup {\mathbb P}^1 ....\sqcup {\mathbb P}^1$ consists of $n$ disjoint copies of ${\mathbb P}^1$s.
\begin{table}[h]
  \centering
 \begin{tabular}{|c|c|c|c|c|c|c|c|c|}
\hline
\cellcolor[gray]{0.9}  &\cellcolor[gray]{0.9} $D_1$  &\cellcolor[gray]{0.9} $D_2$  &\cellcolor[gray]{0.9} $D_3$  & \cellcolor[gray]{0.9}$D_4$  & \cellcolor[gray]{0.9}$D_5$ &\cellcolor[gray]{0.9} $D_6$  & \cellcolor[gray]{0.9}$D_7$ & \cellcolor[gray]{0.9}$D_8$ \\
    \hline
		\hline
\cellcolor[gray]{0.9}$D_1$ & $-$  &  $\emptyset$      &  ${\mathbb T}^2$        &  $\emptyset$   &  ${\mathbb T}^2$  &  ${\mathbb T}^2$   &  ${\cal C}_3$ & $\emptyset$ \\
\cellcolor[gray]{0.9}$D_2$ & $\emptyset$  &  $-$      &  ${\mathbb T}^2$        &  $\emptyset$   &  ${\mathbb T}^2$  &  ${\mathbb T}^2$   &  $\emptyset$ & ${\cal C}_3$
\\
\cellcolor[gray]{0.9}$D_3$  & ${\mathbb T}^2$  &  ${\mathbb T}^2$      &  ${\cal C}_5$        &  ${\cal C}_3$   &  ${\cal C}_{5}$  &  $2{\mathbb P}^1$   &  $4{\mathbb P}^1$ & $4{\mathbb P}^1$ \\
\cellcolor[gray]{0.9}$D_4$  & $\emptyset$  &  $\emptyset$      &  ${\cal C}_3$        &  $\emptyset$   &  ${\cal C}_{3}$  &  $\emptyset$   &  ${\cal C}_3$ & ${\cal C}_3$ \\
\cellcolor[gray]{0.9}$D_5$ & ${\mathbb T}^2$  &  ${\mathbb T}^2$      &  ${\cal C}_{5}$        &  ${\cal C}_{3}$   &  ${\cal C}_{5}$  &  $2{\mathbb P}^1$   &  $4{\mathbb P}^1$ & $4{\mathbb P}^1$ \\
\cellcolor[gray]{0.9}$D_6$ & ${\mathbb T}^2$  &  ${\mathbb T}^2$      &  $2{\mathbb P}^1$        &  $\emptyset$  &  $2{\mathbb P}^1$  &  $-$   &  $4{\mathbb P}^1$ & $4{\mathbb P}^1$ \\
\cellcolor[gray]{0.9}$D_7$ & ${\cal C}_3$  &  $\emptyset$      &  $4{\mathbb P}^1$        &  ${\cal C}_3$   &  $4{\mathbb P}^1$  &  $4{\mathbb P}^1$   &  $20{\mathbb P}^1$ & $4{\mathbb P}^1$ \\
\cellcolor[gray]{0.9}$D_8$ & $\emptyset$  &  ${\cal C}_3$      &  $4{\mathbb P}^1$        &  ${\cal C}_3$   &  $4{\mathbb P}^1$  &  $4{\mathbb P}^1$   &  $4{\mathbb P}^1$ & $20 {\mathbb P}^1$ \\
\hline
  \end{tabular}
  \caption{Intersection curves of the two coordinate divisors.}
\label{Tab3}
\end{table}

\noindent
Considering the basis of smooth divisors $\{D_1, D_2, D_4, D_7\}$ we get the following triple intersection polynomial:
\bea
\label{eq:I3-1}
& & I_3 = 2D_1^3 + 2D_2^3 - 4 D_1^2 D_7 + 8 D_1 D_7^2 + 4 D_4 D_7^2 - 20 D_7^3,
\eea
and considering the K\"ahler form $J = r^1 D_1+r^2 D_2+r^4 D_4+r^7 D_7$ where $r^i$'s are two-cycle volumes, the overall volume ${\cal V}$ of the CY threefold can given as follows,
\bea
\label{eq:CYvol1}
& & {\cal V} = \frac{1}{6}\,(r^1)^3 + \frac{1}{6}\, (r^2)^3 - 2(r^1)^2 (r^7) + 4\, (r^1) (r^7)^2 + 2(r^4) (r^7)^2 - \frac{10}{3}\, (r^7)^3.
\eea
Further, using the Hodge diamonds of the eight toric coordinate divisors in Eq.~(\ref{eq:HodgeDiamond1}) and looking at the intersection number constraints for del-Pezzo surfaces as given in Eq.~(\ref{eq:dP}), we find that the divisors $D_1$ and $D_2$ are dP$_7$ surfaces which also turn out to be `diagonal' as defined in Eq.~(\ref{eq:diagdP}). Further, $D_4$ is a K3 surface while the divisor $D_6$ is a Wilson divisor which may be suitable for poly-instanton inflation if the orientifold involution results in $h^{0,1}(W) = 1 = h^{0,1}_+(W)$. Further, the second Chern-class of the CY is given by,
\bea
\label{eq:CYc2}
c_2({\rm CY}) = 4 D_6^2 + 26 D_4 D_8 - 12 D_6 D_8 + 6 D_8^2,
\eea
and subsequently, using the second Chern class, Euler characteristics of the divisors and the respective second Chern numbers are given as,
\bea
\label{eq:Dpi}
& & \chi(D_\alpha) = \int_{\rm CY} c_2({\rm CY}) \wedge \hat{D}_\alpha + \hat{D}_\alpha^3 = \{10, 10, 44, 24, 44, 4, -4, -4\}, \\
& & \Pi(D_\alpha) = \int_{\rm CY} c_2({\rm CY}) \wedge \hat{D}_\alpha = \{8, 8, 40, 24, 40, 8, 16, 16\} \quad {\rm for} \, \alpha \in \{1, 2,..,8\}. \nonumber
\eea
Further, the K\"ahler cone for this setup turns out to be of a non-simplicial nature, and it is described by the following conditions on the two-cycle volumes,
\bea
\label{eq:KC1}
& \hskip-1cm {\rm KC:}& \quad  -r^1 + 2 \,r^7> 0\,, \quad -r^2 > 0\,, \quad r^4 - r^7 > 0\,, \quad r^1 - r^7 > 0\,, \quad r^2 + r^7 > 0\,.
\eea

\subsubsection*{Diagonal basis of divisors}
Next, we consider a better basis of smooth divisors in which the CY volume expression takes a simpler form, and the diagonality of the two del-Pezzo divisors in manifested. It turns out that the following divisor basis $\{\Gamma_\alpha\}$ does the needful:
\bea
\label{eq:DiagonalBasis}
& & \Gamma_1 = D_1, \quad \Gamma_2 = D_2, \quad \Gamma_3 = 3 D_7 + D_4 + 6 D_1, \quad \Gamma_4 = D_4,
\eea
where $\Gamma_3$ is a smooth divisor with $\chi(\Gamma_3) = 120 = \Pi(\Gamma_3)$ and having the following Hodge diamond,
\bea
& & \Gamma_3 \equiv \begin{tabular}{ccccc}
    & & 1 & & \\
   & 4 & & 4 & \\
  13 & & 108 & & 13 \\
   & 4 & & 4 & \\
    & & 1 & & \\
  \end{tabular}. \nonumber
\eea
Now the intersection polynomial is simply given as,
\bea
\label{eq:I3}
& & I_3 = 2\Gamma_1^3 + 2\Gamma_2^3 + 36 \, \Gamma_3^2\,\Gamma_4.
\eea
Now, considering the K\"ahler form $J = \sum\limits_{\alpha =1}^4 t^\alpha \Gamma_\alpha$, the overall volume ${\cal V}$ of the CY threefold takes the simple following form,
\bea
\label{eq:CYvol2}
& & {\cal V} = \frac{1}{3} (t^1)^3 + \frac{1}{3} (t^2)^3 + 18\, (t^4) (t^3)^2\,.
\eea
The 4-cycle volume moduli, $\tau_\alpha = \partial_\alpha {\cal V}$, are given as below
\bea
\label{eq:tau's}
& & \hskip-1cm \tau_1 = (t^1)^2,  \qquad  \tau_2 = (t^2)^2, \quad \tau_3 = 36 (t^3) (t^4), \qquad \tau_4  \, = 18\, (t^3)^2, \nonumber
\label{Taus}
\eea
which satisfy the general relation: $t^\alpha \tau_\alpha = 3 {\cal V}$. Moreover, the K\"ahler cone constraints in the new basis are given as below,
\bea
\label{eq:KC2}
& \hskip-1cm {\rm KC:}& \quad  -t^1 > 0\,, \quad -t^2 > 0\,, \quad t^4 - 2 t^3 > 0\,, \quad t^1 + 3 t^3, \quad t^2 + 3 t^3 > 0,
\eea
where the first two conditions corresponds to the exceptional divisors, and dictates the negative Swiss-Cheese like contributions in the overall volume of the CY threefolds. This is given as below 
\bea
\label{eq:CYvol}
& & {\cal V} = \frac{1}{6\sqrt2} \tau_3 \sqrt{\tau_4} -\frac{1}{3} \tau_1^{3/2} - \frac{1}{3} \tau_2^{3/2} \,,
\eea
which clearly shows the well known typical appearance of K3 fibre volume as $\sqrt{\tau_4}$ while the two diagonal del-Pezzo contributing negatively to the swiss-cheese volume. Moreover, the K\"ahler cone conditions (\ref{eq:KC2}) translates into the following conditions,
\bea
\label{eq:KC3}
& & 0 < 2 \tau_1 < \tau_4, \quad 0 < 2 \tau_2 < \tau_4, \quad \tau_4 < \left(\frac{3}{\sqrt2}{\cal V} + \frac{1}{\sqrt2}(\tau_1^{3/2}+\tau_2^{3/2})\right)^{2/3} \simeq {\cal V}^{2/3}.
\eea
This shows that for a fixed CY volume, the K3-fibre modulus $\tau_4$ is bounded, and it has been found earlier that such a  constraint is quite strong on the inflaton field range in the corresponding Fibre inflation models \cite{Cicoli:2018tcq}.

Subsequently, using the K\"ahler form in the diagonal divisor basis $J = t^1 \Gamma_1 + t^2 \Gamma_2 +t^3 \Gamma_3 + t^4 \Gamma_4$ the respective sizes of the curves lying at the intersections of the two toric coordinates divisors can be computed as,
\bea
& & \int_{D_\beta \cap D_\gamma} J = \int_{\rm CY} J \wedge \hat{D}_\beta \wedge \hat{D}_\gamma = t^\alpha \int_{\rm CY} \hat{\Gamma}_\alpha \wedge \hat{D}_\beta \wedge \hat{D}_\gamma,
\eea
which are presented in Table \ref{Tab4} where $\hat{t} = 2(t^1+t^2+8 t^3+2 t^4)$. Comparing the two Mori cones we find that they are the same and hence the K\"ahler cone conditions genuinely corresponds to the actual CY threefolds \cite{Cicoli:2018tcq}. Also, this table is symmetrical and lower left entries can be read-off from the upper right block.
\begin{table}[h]
  \centering
 \begin{tabular}{|c|c|c|c|c|c|c|c|c|}
\hline
 \cellcolor[gray]{0.9} &\cellcolor[gray]{0.9} $D_1$  &\cellcolor[gray]{0.9} $D_2$  & \cellcolor[gray]{0.9}$D_3$  &\cellcolor[gray]{0.9} $D_4$  &\cellcolor[gray]{0.9} $D_5$ & \cellcolor[gray]{0.9}$D_6$  &\cellcolor[gray]{0.9} $D_7$ &\cellcolor[gray]{0.9} $D_8$ \\
    \hline
		\hline
\cellcolor[gray]{0.9}$D_1$ & $2t^1$ & 0 & $-2t^1$ & 0 & $-2t^1$ & $-2t^1$ & $-4 t^1$ & 0 \\
\cellcolor[gray]{0.9}$D_2$ &  & $2t^2$ & $-2t^2$ & 0 & $-2t^2$ & $-2t^2$ & 0 & $-4t^2$ \\
\cellcolor[gray]{0.9}$D_3$ &  &  & $\hat{t}$ & $12t^3$ & $\hat{t}$ & $2(t^1+t^2+6 t^3)$ & $4(t^1+t^3+t^4)$ & $4(t^2+t^3+t^4)$\\
\cellcolor[gray]{0.9}$D_4$ &  &  &  & 0 & $12t^3$ & 0 & $12t^3$ & $12t^3$\\
\cellcolor[gray]{0.9}$D_5$  & &  &  &  & $\hat{t}$ & $2(t^1+t^2+6t^3)$ & $4(t^1+4 t^3+4t^4)$ & $4(t^2+t^3+t^4)$ \\
\cellcolor[gray]{0.9}$D_6$  &  &  &  &  &  & $2(t^1+t^2)$ & $4(t^1+3t^3)$ & $4(t^2+3t^3)$\\
\cellcolor[gray]{0.9}$D_7$  &  &  &  &  &  &  & $4(2t^1-2t^3+t^4)$ & $4(t^4-2t^3)$ \\
\cellcolor[gray]{0.9}$D_8$  &  &  &  &  &  &  &  & $4(2t^2-2t^3+t^4)$ \\
\hline
  \end{tabular}
  \caption{Size of curves at the intersection locus of two  divisors presented in Table \ref{Tab3}.}
\label{Tab4}
\end{table}

\noindent
Given that the second Chern numbers for the new divisor basis are $\Pi(\Gamma_\alpha) = \{8, 8, 120, 24\}$, one has the following form of the higher derivative $F^4$-corrections to the scalar potential,
\bea
\label{eq:F^4-term-globalmodel}
& & \hskip-1cm V_{{\rm F}^4} = - \frac{\lambda\,\kappa^2\,|W_0|^4}{g_s^{3/2} {\cal V}^4} \, \left(8 t^1 + 8 t^2 + 120 t^3 + 24 t^4\right)\\
& & \hskip-0.3cm = - \frac{2\sqrt2\,\lambda\,\kappa^2\,|W_0|^4}{g_s^{3/2} {\cal V}^4} \, \left(10\sqrt{\tau_4} + \frac{\tau_3}{\sqrt{\tau_4}} -2\sqrt{2} \sqrt{\tau_1} - 2\sqrt{2} \sqrt{\tau_2}\right).\nonumber
\eea


\subsubsection*{Orientifold involution, fluxes and brane setting}
For a given holomorphic involution, one needs to introduce D3/D7-branes and fluxes in order to cancel all the charges. For example, one can nullify the D7-tadpoles via introducing stacks of $N_a$ D7-branes wrapped around suitable divisors (say $D_a$) and their orientifold images ($D_a^\prime$) such that the following relation holds \cite{Blumenhagen:2008zz}:
\bea
\label{eq:D7tadpole}
& & \sum_a\, N_a \left([D_a] + [D_a^\prime] \right) = 8\, [{\rm O7}]\,.
\eea
Moreover, the presence of D7-branes and O7-planes also contributes to the D3-tadpoles, which, in addition, receive contributions from  $H_3$ and $F_3$ fluxes, D3-branes and O3-planes. The D3-tadpole cancellation condition is given as \cite{Blumenhagen:2008zz}:
\be
N_{\rm D3} + \frac{N_{\rm flux}}{2} + N_{\rm gauge} = \frac{N_{\rm O3}}{4} + \frac{\chi({\rm O7})}{12} + \sum_a\, \frac{N_a \left(\chi(D_a) + \chi(D_a^\prime) \right) }{48}\,,
\label{eq:D3tadpole}
\ee
where $N_{\rm flux} = (2\pi)^{-4} \, (\alpha^\prime)^{-2}\int_X H_3 \wedge F_3$ is the contribution from background fluxes and $N_{\rm gauge} = -\sum_a (8 \pi)^{-2} \int_{D_a}\, {\rm tr}\, {\cal F}_a^2$ is due to D7 worldvolume fluxes. However, for the simple case where D7-tadpoles are canceled by placing 4 D7-branes (plus their images) on top of an O7-plane, (\ref{eq:D3tadpole}) reduces to the following form:
\bea
\label{eq:D3tadpole1}
& & N_{\rm D3} + \frac{N_{\rm flux}}{2} + N_{\rm gauge} =\frac{N_{\rm O3}}{4} + \frac{\chi({\rm O7})}{4}\, \in {\mathbb N}.
\eea
For this CY threefold, we note that there are eight reflection involutions $\sigma_i: x_i \to - x_i$ corresponding to flipping each of the eight coordinates, for each $i \in \{1, 2, .., 8\}$. The details of the respective Fixed points are summarized in Table \ref{tab_FixedPointSet}.

\begin{table}[H]
\centering
\hskip0.11cm \begin{tabular}{|c||c|c|c|}
\hline
Inv. & O7-planes & O3-planes  & $Q_{D3}$ \\
\hline
$\sigma_1$ & $\{D_1, D_2, D_4\}$ & $\{D_6 D_7 D_8\}: 4$ & 24 \\
$\sigma_2$ & $\{D_1, D_2, D_4\}$ & $\{D_6 D_7 D_8\}: 4$ & 24 \\
$\sigma_3$ & $\{D_3\}$ & $\{D_1 D_5 D_6, D_2 D_5 D_6\}$: 2+2 & 24 \\
$\sigma_4$ & $\{D_1, D_2, D_4\}$ & $\{D_6 D_7 D_8\}: 4$ & 24 \\
$\sigma_5$ & $\{D_5\}$ & $\{D_1 D_3 D_6, D_2 D_3 D_6\}: 2+2$ & 76 \\
$\sigma_6$ & $\{D_6\}$ & $\{D_1 D_3 D_5, D_2 D_3 D_5, D_3 D_4 D_5, D_4 D_7 D_8\}: 2+2+4+4$ & 8 \\
$\sigma_7$ & $\{D_7\}$ & $\{D_2 D_6 D_8\}: 2$ & 0 \\
$\sigma_8$ & $\{D_8\}$ & $\{D_1 D_6 D_7\}: 2$ & 0 \\
\hline
\end{tabular}
\caption{Fixed point set for a given involution $\sigma_i: x_i \to - x_i$. Here, the D3 tadpole charge $Q_{\rm D3}$ corresponds to the brane setting where D7-branes are placed on top of the O7-planes.}
\label{tab_FixedPointSet}
\end{table}


\subsection{Example 2}
The CY threefold corresponding to the polytope Id: 686 is defined by the following toric data:
\begin{center}
\begin{tabular}{|c|cccccccc|}
\hline
\cellcolor[gray]{0.9}Hyp &\cellcolor[gray]{0.9} $x_1$  &\cellcolor[gray]{0.9} $x_2$  &\cellcolor[gray]{0.9} $x_3$  &\cellcolor[gray]{0.9} $x_4$  &\cellcolor[gray]{0.9} $x_5$ & \cellcolor[gray]{0.9}$x_6$  &\cellcolor[gray]{0.9} $x_7$ &\cellcolor[gray]{0.9} $x_8$      \\
\hline
\cellcolor[gray]{0.9} 6 & 0 & 0 & 1 & 0 & 3 & 0 & 1 & 1  \\
\cellcolor[gray]{0.9} 6 & 0 & 0 & 1 & 1 & 3 & 1 & 0 & 0 \\
\cellcolor[gray]{0.9} 12 & 1 & 0 & 2 & 1 & 6 & 0 & 0 & 2 \\
\cellcolor[gray]{0.9} 12 & 1 & 1 & 2 & 2 & 6 & 0  & 0 & 0 \\
\hline
& dP$_8$  & dP$_8$ &  &  K3 &  & $D_W$ &  &   \\
\hline
\end{tabular}
\end{center}
The Hodge numbers are $(h^{2,1}, h^{1,1}) = (70, 4)$, the Euler number is $\chi=-132$ and the Stanley-Reisner ideal is:
\be
{\rm SR} =  \{x_1 x_2, x_1 x_4, x_1 x_8, x_2 x_4, x_2 x_7, x_4 x_6, x_3 x_5 x_6 x_7, x_3 x_5 x_6 x_8, x_3 x_5 x_7 x_8\} \,. \nn
\ee
The analysis of the divisor topologies using {\it cohomCalg} \cite{Blumenhagen:2010pv, Blumenhagen:2011xn} shows that they can be represented by the following Hodge diamonds:
\bea
\label{eq:HodgeDiamond2}
D_1 &\equiv& \begin{tabular}{ccccc}
    & & 1 & & \\
   & 0 & & 0 & \\
  0 & & 9 & & 0\\
   & 0 & & 0 & \\
    & & 1 & & \\
  \end{tabular} \equiv D_2, \qquad \quad \, \,  D_3 \equiv \begin{tabular}{ccccc}
    & & 1 & & \\
   & 0 & & 0 & \\
  2 & & 28 & & 2 \\
   & 0 & & 0 & \\
    & & 1 & & \\
  \end{tabular}, \\
& & \nonumber\\
D_4 &\equiv& \begin{tabular}{ccccc}
    & & 1 & & \\
   & 0 & & 0 & \\
  1 & & 20 & & 1 \\
   & 0 & & 0 & \\
    & & 1 & & \\
  \end{tabular} \equiv K3, \qquad \quad D_5 \equiv \begin{tabular}{ccccc}
    & & 1 & & \\
   & 0 & & 0 & \\
  16 & & 116 & & 16 \\
   & 0 & & 0 & \\
    & & 1 & & \\
  \end{tabular}, \nonumber\\
& & \nonumber\\
D_6 &\equiv& \begin{tabular}{ccccc}
    & & 1 & & \\
   & 1 & & 1 & \\
  0 & & 4 & & 0 \\
   & 1 & & 1 & \\
    & & 1 & & \\
  \end{tabular} \equiv D_W, \qquad \quad \, \, D_7 \equiv \begin{tabular}{ccccc}
    & & 1 & & \\
   & 2 & & 2 & \\
  0 & & 4 & & 0 \\
   & 2 & & 2 & \\
    & & 1 & & \\
  \end{tabular} \equiv D_8. \nonumber
\eea
Let us make an important observation that this CY threefold has an exchange symmetry given by $\{1 \leftrightarrow 2, 7 \leftrightarrow 8\}$, which is present at the level of SR ideal and the Hodge Diamonds of the corresponding toric divisors. The intersection curves lying at the two toric coordinate divisors are presented in Table \ref{Tab3} where ${\cal C}_g$ denotes a curve of genus $g$ which is given by the Hodge number $\{h^{0,0} = 1, h^{1,0} = g \}$, i.e. ${\cal C}_0 = {\mathbb P}^1, {\cal C}_1 = {\mathbb T}^2$, and $n {\mathbb P}^1 = {\mathbb P}^1 \sqcup {\mathbb P}^1 ....\sqcup {\mathbb P}^1$ consists of $n$ disjoint copies of ${\mathbb P}^1$s.
\begin{table}[h]
  \centering
 \begin{tabular}{|c|c|c|c|c|c|c|c|c|}
\hline
\cellcolor[gray]{0.9}  &\cellcolor[gray]{0.9} $D_1$  &\cellcolor[gray]{0.9} $D_2$  &\cellcolor[gray]{0.9} $D_3$  & \cellcolor[gray]{0.9}$D_4$  & \cellcolor[gray]{0.9}$D_5$ &\cellcolor[gray]{0.9} $D_6$  & \cellcolor[gray]{0.9}$D_7$ & \cellcolor[gray]{0.9}$D_8$ \\
    \hline
		\hline
\cellcolor[gray]{0.9}$D_1$ & $-$  &  $\emptyset$      &  ${\mathbb T}^2$        &  $\emptyset$   &  ${\cal C}_4$  &  ${\mathbb T}^2$   &  ${\cal C}_2$ & $\emptyset$ \\
\cellcolor[gray]{0.9}$D_2$ & $\emptyset$  &  $-$      &  ${\mathbb T}^2$        &  $\emptyset$   &  ${\cal C}_4$  &  ${\mathbb T}^2$   &  $\emptyset$ & ${\cal C}_2$
\\
\cellcolor[gray]{0.9}$D_3$  & ${\mathbb T}^2$  &  ${\mathbb T}^2$      &  ${\cal C}_3$        &  ${\cal C}_2$   &  ${\cal C}_{13}$  &  ${\mathbb P}^1$   &  $2{\mathbb P}^1$ & $2{\mathbb P}^1$ \\
\cellcolor[gray]{0.9}$D_4$  & $\emptyset$  &  $\emptyset$      &  ${\cal C}_2$        &  $\emptyset$   &  ${\cal C}_{10}$  &  $\emptyset$   &  ${\cal C}_2$ & ${\cal C}_2$ \\
\cellcolor[gray]{0.9}$D_5$ & ${\cal C}_4$  &  ${\cal C}_4$      &  ${\cal C}_{13}$        &  ${\cal C}_{10}$   &  ${\cal C}_{55}$  &  $3{\mathbb P}^1$   &  $6{\mathbb P}^1$ & $6{\mathbb P}^1$ \\
\cellcolor[gray]{0.9}$D_6$ & ${\mathbb T}^2$  &  ${\mathbb T}^2$      &  ${\mathbb P}^1$        &  $\emptyset$   &  $3{\mathbb P}^1$  &  $-$   &  $2{\mathbb P}^1$ & $2{\mathbb P}^1$ \\
\cellcolor[gray]{0.9}$D_7$ & ${\cal C}_2$  &  $\emptyset$      &  $2{\mathbb P}^1$        &  ${\cal C}_2$   &  $6{\mathbb P}^1$  &  $2{\mathbb P}^1$   &  $10{\mathbb P}^1$ & $2{\mathbb P}^1$ \\
\cellcolor[gray]{0.9}$D_8$ & $\emptyset$  &  ${\cal C}_2$      &  $2{\mathbb P}^1$        &  ${\cal C}_2$   &  $6{\mathbb P}^1$  &  $2{\mathbb P}^1$   &  $2{\mathbb P}^1$ & $10 {\mathbb P}^1$ \\
\hline
  \end{tabular}
  \caption{Intersection curves of the two coordinate divisors.}
\label{Tab3-ex2}
\end{table}

\noindent
Considering the basis of smooth divisors $\{D_1, D_2, D_4, D_7\}$ we get the following triple intersection polynomial:
\bea
\label{eq:I3-1}
& & I_3 = D_1^3 + D_2^3 - 2 D_1^2 D_7 + 4 D_1 D_7^2 + 2 D_4 D_7^2 - 10 D_7^3,
\eea
and considering the K\"ahler form $J = r^1 D_1+r^2 D_2+r^4 D_4+r^7 D_7$ where $r^i$'s are two-cycle volumes, the overall volume ${\cal V}$ of the CY threefold can given as follows,
\bea
\label{eq:CYvol1}
& & {\cal V} = \frac{1}{6}\,(r^1)^3 + \frac{1}{6}\, (r^2)^3 - (r^1)^2 (r^7) + 2\, (r^1) (r^7)^2 + (r^4) (r^7)^2 - \frac{5}{3}\, (r^7)^3.
\eea
Further, using the Hodge diamonds of the eight toric coordinate divisors in Eq.~(\ref{eq:HodgeDiamond2}) and looking at the intersection number constraints for del-Pezzo surfaces as given in Eq.~(\ref{eq:dP}), we find that the divisors $D_1$ and $D_2$ are dP$_8$ surfaces which also turn out to be `diagonal' as defined in Eq.~(\ref{eq:diagdP}). Further, $D_4$ is a K3 surface while the divisor $D_6$ is a Wilson divisor which may be suitable for poly-instanton inflation if the orientifold involution results in $h^{0,1}(W) = 1 = h^{0,1}_+(W)$. Further, the second Chern-class of the CY is given by,
\bea
\label{eq:CYc2}
c_2({\rm CY}) = 10 D_6^2 + 50 D_4 D_8 - 24 D_6 D_8 + 12 D_8^2,
\eea
and subsequently, using the second Chern class, Euler characteristics of the divisors and the respective second Chern numbers are given as,
\bea
\label{eq:Dpi}
& & \chi(D_\alpha) = \int_{\rm CY} c_2({\rm CY}) \wedge \hat{D}_\alpha + \hat{D}_\alpha^3 = \{11, 11, 34, 24, 150, 2, -2, -2\}, \\
& & \Pi(D_\alpha) = \int_{\rm CY} c_2({\rm CY}) \wedge \hat{D}_\alpha = \{10, 10, 32, 24, 96, 4, 8, 8\} \quad {\rm for} \, \alpha \in \{1, 2,..,8\}. \nonumber
\eea
Further, the K\"ahler cone for this setup turns out to be of a non-simplicial nature, and it is described by the following conditions on the two-cycle volumes,
\bea
\label{eq:KC1}
& \hskip-1cm {\rm KC:}& \quad  -r^1 + 2 \,r^7> 0\,, \quad -r^2 > 0\,, \quad r^4 - r^7 > 0\,, \quad r^1 - r^7 > 0\,, \quad r^2 + r^7 > 0\,.
\eea

\subsubsection*{Diagonal basis of divisors}
Next, we consider a better basis of smooth divisors in which the CY volume expression takes a simpler form, and the diagonality of the two del-Pezzo divisors in manifested. It turns out that the following divisor basis $\{\Gamma_\alpha\}$ does the needful:
\bea
\label{eq:DiagonalBasis}
& & \Gamma_1 = D_1, \quad \Gamma_2 = D_2, \quad \Gamma_3 = 3 D_7 + D_4 + 6 D_1, \quad \Gamma_4 = D_4,
\eea
where $\Gamma_3$ is a smooth divisor with $\chi(\Gamma_3) = 108 = \Pi(\Gamma_3)$ and having the following Hodge diamond,
\bea
& & \Gamma_3 \equiv \begin{tabular}{ccccc}
    & & 1 & & \\
   & 2 & & 2 & \\
  10 & & 94 & & 10 \\
   & 2 & & 2 & \\
    & & 1 & & \\
  \end{tabular}. \nonumber
\eea
Now the intersection polynomial is simply given as,
\bea
\label{eq:I3}
& & I_3 = \Gamma_1^3 + \Gamma_2^3 + 18 \, \Gamma_3^2\,\Gamma_4.
\eea
The eight toric divisors $D_\alpha$ are related to the diagonal basis divisors $\Gamma_\alpha$ as,
\bea
& & \hskip-1.5cm D_1 = \Gamma_1, \quad D_2 = \Gamma_2, \quad D_3 = \frac{1}{3}\left(\Gamma_3 + 2 \Gamma_4 - 3 \Gamma_1 - 3 \Gamma_2 \right), \quad D_4 = \Gamma_4, \quad D_5 = 3 \Gamma_3, \\
& & \hskip-1.5cm D_6 = \Gamma_4 - \Gamma_1 -\Gamma_2, \quad D_7 = \frac{1}{3}\left(\Gamma_3 - \Gamma_4 - 6 \Gamma_1 \right), \quad D_8 = \frac{1}{3}\left(\Gamma_3 - \Gamma_4 - 6 \Gamma_2 \right). \nonumber
\eea
Now, considering the K\"ahler form $J = \sum\limits_{\alpha =1}^4 t^\alpha \Gamma_\alpha$, the overall volume ${\cal V}$ of the CY threefold given in Eq.~(\ref{eq:CYvol1}) takes the simple following form,
\bea
\label{eq:CYvol2}
& & {\cal V} = \frac{1}{6} (t^1)^3 + \frac{1}{6} (t^2)^3 + 9\, (t^4) (t^3)^2\,.
\eea
The 4-cycle volume moduli, $\tau_\alpha = \partial_\alpha {\cal V}$, are given as below
\bea
\label{eq:tau's}
& & \hskip-1cm \tau_1 = \frac{1}{2} (t^1)^2,  \qquad  \tau_2 = \frac{1}{2} (t^2)^2, \quad \tau_3 = 18 (t^3) (t^4), \qquad \tau_4  \, = 9\, (t^3)^2, \nonumber
\label{Taus}
\eea
which satisfy the general relation: $t^\alpha \tau_\alpha = 3 {\cal V}$. Moreover, the K\"ahler cone constraints (\ref{eq:KC1}) in the new basis (\ref{eq:DiagonalBasis}) are given as below,
\bea
\label{eq:KC2}
& \hskip-1cm {\rm KC:}& \quad  -t^1 > 0\,, \quad -t^2 > 0\,, \quad t^4 - 2 t^3 > 0\,, \quad t^1 + 3 t^3, \quad t^2 + 3 t^3 > 0,
\eea
where the first two conditions corresponds to the exceptional divisors, and didctates the negative swiss-cheese like contributions in the overall volume of the CY threefolds. This is given as below 
\bea
\label{eq:CYvol}
& & {\cal V} = \frac{1}{6} \tau_3 \sqrt{\tau_4} -\frac{\sqrt2}{3} \tau_1^{3/2} - \frac{\sqrt2}{3} \tau_2^{3/2} \,,
\eea
which clearly shows the well known typical appearance of K3 fibre volume as $\sqrt{\tau_4}$ while the two diagonal del-Pezzo contributing negatively to the swiss-cheese volume. Moreover, the K\"ahler cone conditions (\ref{eq:KC2}) translates into the following conditions,
\bea
\label{eq:KC3}
& & 0 < 2 \tau_1 < \tau_4, \quad 0 < 2 \tau_2 < \tau_4, \quad \tau_4 < \left(\frac{3}{2}{\cal V} + \frac{1}{\sqrt2}(\tau_1^{3/2}+\tau_2^{3/2})\right)^{2/3} \simeq {\cal V}^{2/3}.
\eea
This shows that for a fixed CY volume, the K3-fibre modulus $\tau_4$ is bounded, and it has been found earlier that such a  constraint is quite strong on the inflaton field range in the corresponding Fibre inflation models \cite{Cicoli:2018tcq}.

Subsequently, using the K\"ahler form in the diagonal divisor basis $J = t^1 \Gamma_1 + t^2 \Gamma_2 +t^3 \Gamma_3 + t^4 \Gamma_4$ the respective sizes of the curves lying at the intersections of the two toric coordinates divisors can be computed as,
\bea
& & \int_{D_\beta \cap D_\gamma} J = \int_{\rm CY} J \wedge \hat{D}_\beta \wedge \hat{D}_\gamma = t^\alpha \int_{\rm CY} \hat{\Gamma}_\alpha \wedge \hat{D}_\beta \wedge \hat{D}_\gamma,
\eea
which are presented in Table \ref{Tab4} where $\tilde{t} = t^1+t^2+8 t^3+2 t^4$. Comparing the two Mori cones we find that they are the same and hence the K\"ahler cone conditions genuinely corresponds to the actual CY threefolds \cite{Cicoli:2018tcq}. Also, this table is symmetrical and lower left entries can be read-off from the upper right block.
\begin{table}[h]
  \centering
 \begin{tabular}{|c|c|c|c|c|c|c|c|c|}
\hline
 \cellcolor[gray]{0.9} &\cellcolor[gray]{0.9} $D_1$  &\cellcolor[gray]{0.9} $D_2$  & \cellcolor[gray]{0.9}$D_3$  &\cellcolor[gray]{0.9} $D_4$  &\cellcolor[gray]{0.9} $D_5$ & \cellcolor[gray]{0.9}$D_6$  &\cellcolor[gray]{0.9} $D_7$ &\cellcolor[gray]{0.9} $D_8$ \\
    \hline
		\hline
\cellcolor[gray]{0.9}$D_1$ & $t^1$ & 0 & $-t^1$ & 0 & $-3 t^1$ & $-t^1$ & $-2 t^1$ & 0 \\
\cellcolor[gray]{0.9}$D_2$ &  & $t^2$ & $-t^2$ & 0 & $-3 t^2$ & $-t^2$ & 0 & $-2 t^2$ \\
\cellcolor[gray]{0.9}$D_3$ &  &  & $\tilde{t}$ & $6 t^3$ & $3\tilde{t}$ & $t^1+t^2+6 t^3$ & $2 t^1+2 t^3+2 t^4$ & $2 t^2+2 t^3+2 t^4$\\
\cellcolor[gray]{0.9}$D_4$ &  &  &  & 0 & $18 t^3$ & 0 & $6 t^3$ & $6 t^3$\\
\cellcolor[gray]{0.9}$D_5$  & &  &  &  & $9 \tilde{t}$ & $3 t^1+3 t^2+18 t^3$ & $6 t^1+6 t^3+6 t^4$ & $6 t^2+6 t^3+6 t^4$ \\
\cellcolor[gray]{0.9}$D_6$  &  &  &  &  &  & $t^1+t^2$ & $2 t^1+6 t^3$ & $2 t^2+6 t^3$\\
\cellcolor[gray]{0.9}$D_7$  &  &  &  &  &  &  & $4 t^1-4 t^3+2 t^4$ & $2 t^4-4 t^3$ \\
\cellcolor[gray]{0.9}$D_8$  &  &  &  &  &  &  &  & $4 t^2-4 t^3+2 t^4$ \\
\hline
  \end{tabular}
  \caption{Size of curves at the intersection locus of two  divisors presented in Table \ref{Tab3-ex2}.}
\label{Tab4-ex2}
\end{table}

\noindent
Given that the second Chern numbers for the new divisor basis are $\Pi(\Gamma_\alpha) = \{10, 10, 24, 108\}$, one has the following form of the higher derivative $F^4$-corrections to the scalar potential,
\bea
\label{eq:F^4-term-globalmodel}
& & \hskip-1cm V_{{\rm F}^4} = - \frac{\lambda\,\kappa^2\,|W_0|^4}{g_s^{3/2} {\cal V}^4} \, \left(10 t^1 + 10 t^2 + 108 t^3 + 24 t^4\right)\\
& & \hskip-0.3cm = - \frac{4\,\lambda\,\kappa^2\,|W_0|^4}{g_s^{3/2} {\cal V}^4} \, \left(9\sqrt{\tau_4} + \frac{\tau_3}{\sqrt{\tau_4}} -\frac{5}{\sqrt{2}} \sqrt{\tau_1} - \frac{5}{\sqrt{2}} \sqrt{\tau_2}\right).\nonumber
\eea


\subsubsection*{Orientifold involution, fluxes and brane setting}
For a given holomorphic involution, one needs to introduce D3/D7-branes and fluxes in order to cancel all the charges. For example, one can nullify the D7-tadpoles via introducing stacks of $N_a$ D7-branes wrapped around suitable divisors (say $D_a$) and their orientifold images ($D_a^\prime$) such that the following relation holds \cite{Blumenhagen:2008zz}:
\bea
\label{eq:D7tadpole}
& & \sum_a\, N_a \left([D_a] + [D_a^\prime] \right) = 8\, [{\rm O7}]\,.
\eea
Moreover, the presence of D7-branes and O7-planes also contributes to the D3-tadpoles, which, in addition, receive contributions from  $H_3$ and $F_3$ fluxes, D3-branes and O3-planes. The D3-tadpole cancellation condition is given as \cite{Blumenhagen:2008zz}:
\be
N_{\rm D3} + \frac{N_{\rm flux}}{2} + N_{\rm gauge} = \frac{N_{\rm O3}}{4} + \frac{\chi({\rm O7})}{12} + \sum_a\, \frac{N_a \left(\chi(D_a) + \chi(D_a^\prime) \right) }{48}\,,
\label{eq:D3tadpole}
\ee
where $N_{\rm flux} = (2\pi)^{-4} \, (\alpha^\prime)^{-2}\int_X H_3 \wedge F_3$ is the contribution from background fluxes and $N_{\rm gauge} = -\sum_a (8 \pi)^{-2} \int_{D_a}\, {\rm tr}\, {\cal F}_a^2$ is due to D7 worldvolume fluxes. However, for the simple case where D7-tadpoles are canceled by placing 4 D7-branes (plus their images) on top of an O7-plane, (\ref{eq:D3tadpole}) reduces to the following form:
\bea
\label{eq:D3tadpole1}
& & N_{\rm D3} + \frac{N_{\rm flux}}{2} + N_{\rm gauge} =\frac{N_{\rm O3}}{4} + \frac{\chi({\rm O7})}{4}\, \in {\mathbb N}.
\eea
For this CY threefold, we note that there are eight reflection involutions $\sigma_i: x_i \to - x_i$ corresponding to flipping each of the eight coordinates, for each $i \in \{1, 2, .., 8\}$. The details of the respective Fixed points are summarized in Table \ref{tab_FixedPointSet}.

\begin{table}[H]
\centering
\hskip0.11cm \begin{tabular}{|c||c|c|c|c|}
\hline
Inv. & O7-planes & O3-planes  & $Q_{D3}$ & $n_B$\\
\hline
$\sigma_1$ & $\{D_1, D_2, D_4\}$ & $\{D_6 D_7 D_8\}: 2$ & 24 & 5\\
$\sigma_2$ & $\{D_1, D_2, D_4\}$ & $\{D_6 D_7 D_8\}: 2$ & 24 & 5\\
$\sigma_3$ & $\{D_3\}$ & $\{D_1 D_5 D_6, D_2 D_5 D_6\}$: 3+3 & 20 & 32 \\
$\sigma_4$ & $\{D_1, D_2, D_4\}$ & $\{D_6 D_7 D_8\}: 2$ & 24 & 5\\
$\sigma_5$ & $\{D_5\}$ & $\{D_1 D_3 D_6, D_2 D_3 D_6\}: 1+1$ & 76 & 697 \\
$\sigma_6$ & $\{D_6\}$ & $\{D_1 D_3 D_5, D_2 D_3 D_5, D_3 D_4 D_5,  $ & 8 & 1\\
& & $D_4 D_7 D_8\} : 3+3+6+2$ & & \\
$\sigma_7$ & $\{D_7\}$ & $\{D_2 D_6 D_8\}: 2$ & 0 & 1\\
$\sigma_8$ & $\{D_8\}$ & $\{D_1 D_6 D_7\}: 2$ & 0 & 1\\
\hline
\end{tabular}
\caption{Fixed point set for a given involution $\sigma_i: x_i \to - x_i$. Here, the D3 tadpole charge $Q_{\rm D3}$ corresponds to the brane setting where D7-branes are placed on top of the O7-planes. In addition, $n_B$ denotes the number of possible brane configurations satisfying the D7-brane tadpole cancellation condition (\ref{eq:D7tadpole}).}
\label{tab_FixedPointSetEx2}
\end{table}


\subsection{Possible contributions to the scalar potential }
There can be several types of (non-)perturbative effects which can induce useful scalar potential contributions, namely the BBHL's $(\alpha^\prime)^3$ corrections \cite{Becker:2002nn}, the non-perturbative effects of \cite{Witten:1996bn} and the higher derivative F$^4$ corrections of \cite{Ciupke:2015msa}. Using the Gukov-Vafa-Witten's (GVW) flux superpotential $W_0$ for stabilizing the complex-structure moduli and the axio-dilaton at their respective supersymmetric minimum, the various scalar potential contributions can be collected in the following form
\bea
\label{eq:masterVgen}
& & \hskip-0.75cm V_{\rm tot} = V_{\rm up} + V_{\rm LVS} + V_{\rm polyinst} + V_{\rm loop} +V_{{\rm F}^4} + \dots 
\eea
where our current goal does not include addressing the de-Sitter uplifting issue, and we assume that any of the well known schemes may be used for the purpose. For example, we consider the three popular classes of uplifting by simply characterizing them via the following contributions to the scalar potential,
\bea
\label{eq:Vuplift}
& & V_{\rm up}({\cal V}) = \frac{{\cal C}_{\rm up}}{{\cal V}^p},
\eea
where $p = 4/3$ for anti-D3 uplifting \cite{Kachru:2003aw,Crino:2020qwk,Cicoli:2017axo,AbdusSalam:2022krp}, and $p = 2$ for D-term uplifting \cite{Burgess:2003ic,Achucarro:2006zf,Braun:2015pza} while  $p = 8/3$ for the T-brane uplifting \cite{Cicoli:2015ylx,Cicoli:2017shd}. 

Let us proceed by giving some further details of the possible corrections to the effective scalar potential. Apart from the standard BBHL corrections, the higher derivative F$^4$-corrections generically contribute to the scalar potential. Also, given that all the reflection involutions $\sigma_i$ result in $O3$-planes in their respective Fixed point sets, KK-type string loop corrections are also generically possible. However, the Winding-type string loop corrections may or may not be contributing depending on the specific choice of the D7-brane setting and the intersection loci of the two D7-stacks. Moreover, the presence of non-perturbative effects is the most tricky part. The E3-instanton or gaugino condensation effects on the D7-branes wrapping the two rigid del-Pezzo divisors are likely to contribute depending on the transverse and longitudinal invariance under the given involution, while the poly-instanton effects are more specific in the sense of having $h^{1,1}_+(D_W) = 1$ for the Wilson divisor $D_W$. The schematic form of the holomorphic superpotential can be given as
\bea
\label{eq:Wnp-pn}
& & \hskip-1.5cm W (T_1, T_2, T_4) 
=  W_0 + \sum_{\alpha = 1}^2 \left(A_{\alpha}\, e^{- i\, a_{\alpha} T_\alpha} + A_\alpha \, A_{w_\alpha}\, e^{- i\, a_{\alpha} T_\alpha} \, e^{-i\, a_w (T_4-T_1-T_2)} \right),
\eea
where $a_{\alpha} = 2\pi/n_\alpha$ for some $m_\alpha \in {\mathbb N}$ depending on the rank of the gauge groups corresponding to the gaugino condensation effects from the D7-brane wrapping the suitable four-cycles, while poly-instanton parameters are given as $a_w = 2 \pi$. In case, the non-perturbative effects arising from the rigid divisor $D_{s_\alpha}$ is induced from the E3-instanton, we have $m_\alpha = 1$. 

\subsubsection*{Leading order LVS contributions}
The LVS piece appearing at the leading order as compared to the poly-instanton and string loop corrections can be given as 
\bea
& & \hskip-1.5cm V_{\rm LVS} \equiv V_{\rm LVS} (\tau_1, \tau_2, {\cal V}, \rho_1, \rho_2) = V_{\alpha^\prime}+V_{\rm LVS}^{\rm (np1)}+ V_{\rm LVS}^{\rm (np2)},
\eea
where $V_{\rm LVS}$ results in the following leading terms in the large volume expansion,
\bea
\label{eq:VlvsSimp}
& & \hskip-1.5cm V_{\rm LVS} \simeq \frac{3\kappa{\hat\xi}|W_0|^2}{4\,{\cal V}^3} + \frac{4 \,\kappa\,W_0}{{\cal V}^2}\sum_{\alpha=1}^2 \tau_\alpha \left(A_\alpha a_\alpha e^{-a_\alpha\tau_\alpha}\, \cos(a_\alpha\rho_\alpha) \right) + \frac{8 \kappa}{3{\cal V}} \sum_{\alpha=1}^2 \, \frac{\sqrt{\tau_\alpha}}{\gamma_\alpha} \biggl[a_\alpha^2 |A_\alpha|^2 e^{-2a_\alpha\tau_\alpha} \biggr].
\eea

\subsubsection*{String-loop corrections}
The effective scalar potential contributions arising from the KK-type and Winding-type string loop corrections generically take the following form,
\bea
& & \hskip-0.91cm V_{\rm loop} \equiv V_{g_s}^{\rm KK} + V_{g_s}^{\rm W} = \frac{\kappa\,|W_0|^2}{{\cal V}^2} \left[\sum_\alpha \left(g_s\, {\cal C}_\alpha^{\rm KK}\right)^2 {\cal G}^{(0)}_{\alpha\alpha} - \sum_\alpha \frac{2\, {\cal C}_\alpha^W}{{\cal V}\, t^\alpha_\cap} \right], 
\eea
where the metric components ${\cal G}^{(0)}_{\alpha\alpha}$ can be read-off from the tree level K\"ahler moduli space metric 
\bea
& & {\cal G}^{(0)}_{\alpha\beta} = \frac{1}{4\,{\cal V}^2}\, \left(2\,t^\alpha t^\beta - 4\, {\cal V} \,k^{\alpha\beta}\right)\,.
\eea
The leading order contributions from such string-loop effects generically appear at ${\cal O}({\cal V}^{-10/3})$ in the large volume expansion. In fact there can be additional loop corrections motivated by the field theoretic computations \cite{vonGersdorff:2005bf,Gao:2022uop}, however, we do not include those corrections in the current analysis.

For the specific class of CY based models presented earlier, we generically have the following types of contributions possible from the KK-type string loop effects,
\bea
\label{eq:VKKLoop}
& & V_{g_s}^{\rm KK} 
\simeq \frac{\kappa\,g_s^2\,|W_0|^2}{{\cal V}^2} \left[\frac{3\gamma_1\, \left({\cal C}_1^{\rm KK}\right)^2}{8{\cal V} \sqrt{\tau_1}} + \frac{3\gamma_2\, \left({\cal C}_2^{\rm KK}\right)^2}{8{\cal V} \sqrt{\tau_2}} + \frac{\gamma_b\, \left({\cal C}_3^{\rm KK}\right)^2\, \tau_4}{2{\cal V}^2} + \frac{\left({\cal C}_4^{\rm KK}\right)^2}{4{\tau_4^2}} \right],
\eea
where explicit components for the leading order K\"ahler moduli space metric are used in arriving at the above expression. Note that after the LVS minimization $\{{\cal V}, \tau_1, \tau_2\}$ are fixed and the $\tau_4$ direction could be inflaton candidate. Further, considering the possible two-cycle volumes at the various intersection loci of two toric divisors as collected in Table \ref{Tab4-ex2}, the possible Winding-type string loop corrections can be schematically given as 
\bea
\label{eq:VWindingLoops}
& & \hskip-1cm V_{g_s}^{\rm W} \simeq -\frac{2\kappa\,|W_0|^2}{{\cal V}^3} \biggl[-\frac{{\cal C}_{1}^W}{t^1}-\frac{{\cal C}_{2}^W}{t^2}+\frac{{\cal C}_{3}^W}{t^3} +\frac{{\cal C}_{4}^W}{t^1+t^2+8t^3+2t^4}\\
& & \hskip0.5cm +\frac{{\cal C}_{5}^W}{t^1+3t^3}+\frac{{\cal C}_{6}^W}{t^2+3t^3} +\frac{{\cal C}_{7}^W}{t^1+t^2+6t^3} +\frac{{\cal C}_{8}^W}{t^1+t^3+t^4}+\frac{{\cal C}_{9}^W}{t^2+t^3+t^4}+\frac{{\cal C}_{10}^W}{t^4-3t^3} \biggr], \nonumber
\eea
where ${\cal C}_{\alpha}^W$'s are various complex structure moduli dependent quantities which can be considered as parameters after the supersymmetric stabilization of these moduli via background fluxes. Moreover, for our explicit CY orientifold examples, many of these corrections could be absent by suitable choice of brane-settings, given that the corresponding divisor intersection loci correspond to a collection of shrinkable ${\mathbb P}^1$ surfaces; for example, Table \ref{Tab3-ex2} shows that the $4\times4$ bottom-right block is either emptyset or a collection of spheres. Subsequently, all the pieces collected in the second line of (\ref{eq:VWindingLoops}) should be absent. Subsequently one has,
\bea
\label{eq:VWindingLoops}
& & \hskip-1cm V_{g_s}^{\rm W} \simeq -\frac{2\kappa\,|W_0|^2}{{\cal V}^3} \biggl[\frac{{\cal C}_{1}^W}{\sqrt2\, \sqrt{\tau_1}}+\frac{{\cal C}_{2}^W}{\sqrt2\, \sqrt{\tau_2}}+\frac{3{\cal C}_{3}^W}{\sqrt{\tau_4}} +\frac{{\cal C}_{4}^W\, \tau_4}{2{\cal V}}\biggr].
\eea
Here, let us emphasize again that the specifics of the string-loop corrections to the effective scalar potential largely depend on the explicit choice of the brane settings, and sometimes it is easily possible to avoid (at least one of) the KK or Winding type corrections by a suitable brane configuration and choosing an appropriate holomorphic involution. However, for fibre inflation purpose one needs `enough' and `appropriate' string loop corrections, and therefore, simply avoiding string-loop corrections does not serve the purpose. At the same time having bad corrections, e.g. those which lead to steepening piece kick early and ruin the inflationary plateau, is also not desirable. These requirements make it a delicate engineering of brane configurations and model dependent parameters. 

\subsubsection*{Higher derivative F$^4$ corrections}
Taking into account the CY geometry presented in Example 2, the so-called higher derivative F$^4$-corrections take the following generic but simple form \cite{Ciupke:2015msa},
\bea
\label{eq:F^4-term-globalmodel}
& & \hskip-1cm V_{{\rm F}^4} = - \frac{\lambda\,\kappa^2\,|W_0|^4}{g_s^{3/2} {\cal V}^4} \, \left(10 t^1 + 10 t^2 + 108 t^3 + 24 t^4\right)\\
& & \hskip-0.3cm = \frac{{\cal C}_\lambda}{{\cal V}^3} \, \left(\frac{6}{\tau_4} + \frac{9\sqrt\tau_4}{{\cal V}} -\frac{5}{\sqrt{2}} \frac{\sqrt{\tau_1}}{{\cal V}} - \frac{5}{\sqrt{2}} \frac{\sqrt{\tau_2}}{{\cal V}}\right); \quad {\cal C}_\lambda = - \frac{4\,\lambda\,\kappa^2\,|W_0|^4}{g_s^{3/2}}.\nonumber
\eea
We note that leading order higher derivative F$^4$ contributions arising from such corrections naively appear at ${\cal O}({\cal V}^{-11/3})$ in the large volume expansion.

\subsubsection*{Poly-instanton corrections}
We recall that poly-instanton contributions always appear on top of a conventional non-perturbative effect either from E3-instanton or gaugino condensations. For that reason, such corrections do not appear with coefficients $|W_0|^2$. They can appear only through a linear $|W_0|$ with single exponential or double exponential. These can be collected as follows
\bea
& & V_{\rm polyinst} = V_{\rm polyinst}^{(1)}+ V_{\rm polyinst}^{(2)},
\eea
where the first piece, in the large volume expansion, results in the following contributions,
\bea
& & \hskip-1.5cm V_{\rm polyinst}^{(1)} \simeq \frac{4\kappa\, |W_0|}{{\cal V}^2} \biggl[\sum_{\alpha=1}^2 \, |A_\alpha| |A_{w_\alpha}| e^{-a_\alpha\tau_\alpha-a_w \tau_w}\, \left(a_\alpha\tau_\alpha + a_{w}\tau_w\right)\, \cos(a_\alpha\rho_\alpha + a_{w_\alpha}\rho_w)\biggr],
\eea
where $\tau_w = \tau_4-\tau_1-\tau_2$ and $\rho_w = \rho_4-\rho_1-\rho_2$. We note that the second piece $V_{\rm polyinst}^{(2)}$ takes a rather too complicated form to illustrate something useful here. However, after LVS minimizing which leaves only $(\tau_4, \rho_4)$ moduli or equivalently $(\tau_w, \rho_w)$ moduli unstabilized, we can write this piece in some simpler form. We recall that $a_w = 2 \pi$ as both correspond to an instanton correction. Subsequently it turns out that the poly-instanton contributions to the effective single-field scalar potential after LVS minimization can be given as below
\bea
& & V_{\rm polyinst} \simeq \left({\cal C}_{\rm poly}^{(1)} + \tau_w {\cal C}_{\rm poly}^{(2)}\right) e^{-2\pi\tau_w}\, \cos(2\pi \rho_w),
\eea
where the coefficients ${\cal C}_{\rm poly}^{(1)}$ and ${\cal C}_{\rm poly}^{(2)}$ are volume suppressed constants after the LVS minimization performed at the leading order. Let us also note that the piece $V_{\rm polyinst}^{(2)}$ also involves the doubly suppressed terms with factors ${\cal O}(e^{-4\pi\tau_w})$ appearing in the poly-instanton series, and one can ignore those in a self-consistent approach of moduli stabilization.

\subsubsection*{Summary}
Finally, the generic scalar potential can be schematically given as below,
\bea
\label{eq:Vfinal-simp}
& & \hskip-0.5cm V \simeq \frac{{\cal C}_{\rm up}}{{\cal V}^p} + \biggl[\frac{3\kappa{\hat\xi}|W_0|^2}{4\,{\cal V}^3} + \frac{4 \,\kappa\,W_0}{{\cal V}^2}\sum_{\alpha=1}^2 \tau_\alpha \left(A_\alpha a_\alpha e^{-a_\alpha\tau_\alpha}\, \cos(a_\alpha\rho_\alpha) \right)\\
& & + \frac{8 \kappa}{3{\cal V}} \sum_{\alpha=1}^2 \, \frac{\sqrt{\tau_\alpha}}{\gamma_\alpha} \left(a_\alpha^2 |A_\alpha|^2 e^{-2a_\alpha\tau_\alpha} \right)\biggr] + \frac{\kappa\,g_s^2\,|W_0|^2}{{\cal V}^2} \biggl[\frac{3\gamma_1\, \left({\cal C}_1^{\rm KK}\right)^2}{8{\cal V} \sqrt{\tau_1}} + \frac{3\gamma_2\, \left({\cal C}_2^{\rm KK}\right)^2}{8{\cal V} \sqrt{\tau_2}} \nonumber\\
& & + \frac{\gamma_b\, \left({\cal C}_3^{\rm KK}\right)^2\, \tau_4}{2{\cal V}^2} + \frac{\left({\cal C}_4^{\rm KK}\right)^2}{4{\tau_4^2}} \biggr] -\frac{2\kappa\,|W_0|^2}{{\cal V}^3} \biggl[\frac{{\cal C}_{1}^W}{\sqrt2\, \sqrt{\tau_1}} +\frac{{\cal C}_{2}^W}{\sqrt2\, \sqrt{\tau_2}} +\frac{3{\cal C}_{3}^W}{\sqrt{\tau_4}} +\frac{{\cal C}_{4}^W\, \tau_4}{2{\cal V}} \biggr] \nonumber\\
& & + \frac{{\cal C}_\lambda}{{\cal V}^3} \, \left(\frac{6}{\tau_4} + \frac{9\sqrt\tau_4}{{\cal V}} -\frac{5}{\sqrt{2}} \frac{\sqrt{\tau_1}}{{\cal V}} - \frac{5}{\sqrt{2}} \frac{\sqrt{\tau_2}}{{\cal V}}\right) \nonumber\\
& &  + \left(\tilde{\cal C}_{\rm poly}^{(1)} + \tau_4 \, \tilde{\cal C}_{\rm poly}^{(2)}\right) e^{-2\pi\tau_4}\, \cos(2\pi \rho_4)   + \dots, \nonumber
\eea
where $\tilde{\cal C}_{\rm poly}^{(1)}$ and $\tilde{\cal C}_{\rm poly}^{(2)}$ are volume dependent coefficients with suppression factor ${\cal V}^{-(3+n)}$ for $n > 0$. As said earlier, the presence of various contributions will depend on the choice of involution and the brane-setting as we will elaborate later on. So, if some brane-setting does not allow a particular correction to be present, one can simply set the corresponding coefficient to zero before performing the moduli stabilization and the inflationary analysis. Let us note the following points about the collection of terms presented in (\ref{eq:Vfinal-simp}):

\begin{itemize}

\item
The first big bracket of the scalar potential (\ref{eq:Vfinal-simp}) denotes the leading order contributions from the $V_{\rm LVS}$ piece which appear at ${\cal O}({\cal V}^{-3})$ in the large volume expansion, along with a suitable uplifting contribution.

\item
The various remaining pieces appear at ${\cal O}({\cal V}^{-(3+n)})$ for $n > 0$ in the large volume expansion. These pieces include the KK-type, Winding-type, F$^4$-type and Poly-instanton type corrections, and can be used to stabilize the LVS flat directions !

\end{itemize}


\section{A unified framework for LVS inflation
}
\label{sec_unifiedLVS}
Now we present a unified framework to realize all the (known) LVS inflationary models based on K\"ahler moduli driven inflationary dynamics. The main idea is that one can realize the respective inflationary potentials of a given type by simply using different orientifolds of the same CY threefold.

\subsection{Suitable orientifolds for (Loop) Blow-up inflation}
\label{sec_BI}
\noindent
For realizing the standard minimal version of the Blow-up inflation \cite{Conlon:2005jm, Bond:2006nc, Cicoli:2017shd}, one needs only two non-perturbative terms along with the BBHL's $\alpha^\prime$ correction; one for realizing LVS and the other one for driving the inflation. 

Any one of the three involutions $\{\sigma_6, \sigma_7, \sigma_8\}$ which corresponds to placing 4 D7-branes (and their image branes) on top of the O7-planes wrapping the divisors $D_6, D_7$ and $D_8$ respectively, will lead to no poly-instanton effects and no string loop corrections of the Winding type. However, KK-type loop corrections may appear due to the presence of O3-planes. In addition, BBHL and F$^4$ corrections will be generically present. We also note that, these brane configurations do not leave space for tuning the $W_0$ parameter which is anyway not a concern in the standard LVS.

Subsequently, the effective scalar potential for this model can be characterized by setting the following conditions in the scalar potential (\ref{eq:Vfinal-simp})
\bea
\label{eq:BI-model1}
& & \tilde{\cal C}_{\rm poly}^{(1)} = 0, \quad \tilde{\cal C}_{\rm poly}^{(2)} = 0, \quad {\cal C}_\alpha^W = 0 \quad \forall \alpha,
\eea
Taking these pieces into account, the simplified version of the effective scalar potential can be given as
\bea
\label{eq:Vfi-simp}
& & \hskip-0.5cm V \simeq V_0(\tau_1,\tau_4,{\cal V}, \rho_1) + V_{\rm BI}(\tau_2,\rho_2).
\eea
Here, the leading piece $V_0$ stabilizing all the saxionic moduli in the LVS minimum while the effective single field potential $V_{\rm BI}$ depends on one of the two blow-up modulus, say $\tau_2$, as below
\bea
\label{eq:LBI}
& & \hskip-1.5cm V_{\rm BI}(\tau_2, \rho_2) \simeq \frac{4 \,\kappa\,W_0}{{\cal V}^2} \tau_2 \left(A_2 a_2 e^{-a_2\tau_2}\, \cos(a_2\rho_2) \right)  + \frac{8 \kappa}{3{\cal V}} \, \frac{\sqrt{\tau_2}}{\gamma_2} \left(a_2^2 |A_2|^2 e^{-2a_2\tau_2}\right) \\
& & + \frac{\kappa\,g_s^2\,|W_0|^2}{{\cal V}^2} \left[\frac{3\gamma_2\, \left({\cal C}_2^{\rm KK}\right)^2}{8{\cal V} \sqrt{\tau_2}} \right] -\frac{2\kappa\,|W_0|^2}{{\cal V}^3} \biggl[\frac{{\cal C}_{2}^W}{\sqrt2\, \sqrt{\tau_2}}\biggr] - \frac{{\cal C}_\lambda}{{\cal V}^3} \, \left(\frac{5}{\sqrt{2}} \frac{\sqrt{\tau_2}}{{\cal V}}\right).\nonumber
\eea
Thus, we have the scalar potential pieces used for the standard Blow-up inflation \cite{Conlon:2005jm, Bond:2006nc, Cicoli:2017shd}. It will be viable subject to justification of stability against the string-loop and F$^4$-corrections. However, our current aim in this work is not to address viability of the LVS inflationary proposal and we limit ourselves to an explict realization of the standard inflationary potentials in explict global constructions.

One of the major challenges for realizing a viable model of blow-up inflation comes from the string loop effects. Although it has been argued that suitable choice of involution can lead to some appropriate brane-setting such that some of string-loop corrections of the KK-type and the Winding-type may be absent. For example, the absence of non-intersecting stacks of D7-brane stacks and O7-planes, along with the absence of O3-planes can ensure the absence of KK-type corrections. Such possibilities have been realized in model proposed in \cite{Cicoli:2017axo, Bera:2024ihl,Leontaris:2025hly}. Similarly, it is also possible that one can have the D7/O7 stacks intersecting on contractible 2-cycles, and such configurations will not induce the Winding-type corrections a la the prescription proposed in \cite{Berg:2004ek, Berg:2005ja, Berg:2005yu, Berg:2007wt, Cicoli:2007xp}; for example see models in \cite{Cicoli:2016xae, Cicoli:2017axo,Bera:2024ihl,Leontaris:2025hly}. However, it is also true that both the KK-type as well as the Winding-type corrections may not be simultaneously avoided, as least we are not aware of such a possibility in a conrete example. Therefore, finding/avoiding the string loop corrections is a very delicate and model dependent task. Having this in mind, it is well anticipated that a `loop blow-up inflation' like proposal is more viable in such settings as the potential (\ref{eq:LBI}) after minimizing the axion $\rho_2$ takes the following form
\bea
& & V(\tau_2) \simeq V_0 + \frac{\tilde{C}_{\rm loop}}{\sqrt\tau_2} + {\tilde{\cal C}_\lambda} \sqrt\tau_2,
\eea
where the exponential pieces being negligibly small are discarded. Such an inflationary potential has been studied in \cite{Bansal:2024uzr}.

\subsection{Suitable orientifolds for Fibre inflation}
For realizing the (standard) fibre inflation setups we take an involution for which poly-instanton effects are identically absent. For that purpose, we find that the relevant involutions to avoid such possibilities are $\{\sigma_1, \sigma_2, \sigma_4\}$. Furthermore, in order to have sufficient and appropriate string loop effects of the KK- and Winding-types, the involutions $\{\sigma_6, \sigma_7, \sigma_8\}$ are not useful, given that they cannot result in a rich enough possibilities for D7-brane setting. Besides, these involutions do not result in Winding-type string loop effects, though KK-type string loop correction may be present for all the reflections as each of them has O3-planes in their respective fixed point sets. Subsequently, it turns out that the most appropriate choices for involutions that have multiple stacks of D7-branes are $\sigma_3$ and $\sigma_5$ only. For this case, a simplified effective scalar potential can be given as
\bea
\label{eq:Vfi-simp}
& & \hskip-0.5cm V \simeq V_0(\tau_1,\tau_2,{\cal V}, \rho_1, \rho_2) + V_{\rm FI}(\tau_4).
\eea
Here, the leading piece $V_0$ stabilizing all the saxionic moduli in the LVS minimum while the effective single field potential $V_{\rm FI}$ depends on the K3 fibre modulus $\tau_4$ as below
\bea
\label{eq:Vfi-simp1}
& & V_{\rm FI}(\tau_4) \simeq {\cal A}_0 + \frac{{\cal A}_1}{{\cal V}^2\tau_4^2} + \frac{{\cal A}_2}{{\cal V}^3\tau_4} + \frac{{\cal A}_3}{{\cal V}^3\sqrt{\tau_4}} + \frac{{\cal A}_4\, \sqrt{\tau_4}}{{\cal V}^4} + \frac{{\cal A}_5\, \tau_4}{{\cal V}^4}
\eea
where ${\cal A}_0$ includes LVS minimization and takes care of the uplifting piece while other coefficients ${\cal A}_i$'s are given as 
\bea
& & {\cal A}_1 = \frac{\kappa}{4}\,g_s^2\,|W_0|^2\left({\cal C}_4^{\rm KK}\right)^2, \qquad {\cal A}_2 = 6{\cal C}_\lambda, \qquad {\cal A}_3 = -6\kappa\, |W_0|^2\, {\cal C}_3^W, \\
& & {\cal A}_4 = 9{\cal C}_\lambda, \qquad {\cal A}_5 = \frac{\kappa}{2}\,|W_0|^2 \left[g_s^2\, \gamma_b\, \left({\cal C}_3^{\rm KK}\right)^2 - 2{\cal C}_4^W\right].\nonumber
\eea
We note that there are a total of three steepening terms one following each from the KK-type, Winding-type and F$^4$-type corrections to the scalar potential. These are collected in the coefficients ${\cal A}_4$ and ${\cal A}_5$. Here we also note that all the steepening terms appears with an extra suppression factor in the large volume limit. Furthermore, the pieces depending on the del-Pezzo divisor volumes are included in the coefficient ${\cal A}_0$.

Now, let us mention the following combinations of scalar potential contributions with the possibilities of brane-settings leading to appropriate  string-loop corrections for fibre inflation:
\begin{itemize}

\item 
{\bf KK-type + Winding-type loop corrections:} 
For standard fibre inflation driven by KK and Winding type string loop corrections \cite{Cicoli:2008gp}, one has to assume/consider that the $\lambda$-parameter governing the F$^4$ correction in small enough \cite{Cicoli:2023njy}. For such setups ${\cal A}_2$ and ${\cal A}_4$ needs to be negligibly small to avoid interfering with the inflationary dynamics.

\item 
{\bf Winding-type loops + F$^4$ corrections:} In some cases, it might be possible to avoid the KK-type loop corrections and have fibre inflation using the Winding-type string loop effects along with the higher derivative F$^4$ corrections. Such models have also been referred to as ``$\alpha^\prime$-inflation" in \cite{Cicoli:2016chb}, and have been presented in \cite{Cicoli:2017axo, Bera:2024ihl, Leontaris:2025hly}. 

\item 
{\bf KK-type loops + F$^4$ corrections:} 
It is possible that the typical Winding-type string-loop effects of BHK prescription can be avoided by a suitable choice of brane-setting. In such cases, all the ${\cal C}$ coefficients corresponding to Winding contributions will be absent, and one would need a combination of the KK-type loop corrections and the higher derivative F$^4$ corrections to drive fibre  inflation.

\item 
{\bf KK-type and Winding-type loops + F$^4$ corrections:}
This can be the most generic case when all the types of corrections are present, however given the steepening contributions having an extra volume suppressed coefficients, it is well anticipated that one would manage to find a suitable parameter space when a viable fibre inflation happens !

\end{itemize}


\subsection{Suitable orientifolds for Poly-instanton inflation}
\noindent
From the two explicit orientifold examples presented in the previous section, we find that only for three (of the eight reflection) involutions corresponding to the toric coordinates, namely $\sigma_1$, $\sigma_2$ and $\sigma_4$, one can have $h^{1,0}(W) = h^{1,0}_+(W/{\cal O}) = 1$ as needed to have the appropriate zero modes in order to induce the poly-instanton effects. In fact, all the three involutions are equivalent in the sense of having the same Fixed point set with 3 components of $O7$-planes wrapping the divisors $\{D1, D_2, D_4\}$, along with a coincident pair of $O3$ planes lying at the point $D_6 D_7 D_8$ of the CY threefold. It is interesting to have three O7-plane components for a setup based on a CY threefold having just $h^{1,1}(\rm CY) = 4$. Further details about fixed-point set are collected in  Table \ref{tab_FixedPointSet}. 

For this set of equivalent involutions, we have five possibilities for the D7-brane configurations that satisfy the tadpole cancelation conditions. These brane configurations are listed as:

\begin{itemize}

\item 
{\bf Model PI$_1$:} Three stacks of 4 D7-branes each (along with their corresponding image branes) wrapping the divisors $\{D_1, D_2, D_4\}$ are placed just on-top of the O7-plane.
\bea
& & 8[O7] = \sum_{\alpha=1, 2, 4}\, 4[D_\alpha] + 4[D_\alpha^\prime],
\eea
which leads to the D3-brane tadpole charge 
\bea
& & \hskip-0.85cm Q_{\rm D_3} = \frac{N_{\rm O3}}{2} + \frac{\chi({\rm O7})}{6} + \sum_\alpha \frac{N_\alpha\left(\chi(D_\alpha)+\chi(D_\alpha^\prime)\right)}{24} \\
& & = \frac{2}{2}+\frac{11+11+24}{6}+\frac{4\cdot 2 \cdot 11+4\cdot 2 \cdot 11+4\cdot 2 \cdot 24}{48} = 24. \nonumber
\eea

\item 
{\bf Model PI$_2$:} 
Four stacks of D7-branes (along with their corresponding image branes) wrapping the divisors $\{D_1, D_2, D_4, D_6\}$ are placed as follows

\bea
& & 8[O7] = \sum_{\alpha=1, 2}\, 5 \cdot \left([D_\alpha] + [D_\alpha^\prime] \right) + 3 \cdot \left([D_4]+ [D_4^\prime]\right) + 1 \cdot\left([D_6]+ [D_6^\prime]\right),
\eea
which leads to the D3-brane tadpole charge 
\bea
& & Q_{\rm D_3} = \frac{2}{2}+\frac{11+11+24}{6}+\frac{5\cdot 2 \cdot 11+5\cdot 2 \cdot 11+3\cdot 2 \cdot 24 + 1 \cdot 2 \cdot 2}{48} = 24.
\eea

\item 
{\bf Model PI$_3$:} 
Four stacks of D7-branes (along with their corresponding image branes) wrapping the divisors $\{D_1, D_2, D_4, D_6\}$ are placed as follows

\bea
& & 8[O7] = \sum_{\alpha=1, 2}\, 6 \cdot \left([D_\alpha] + [D_\alpha^\prime] \right) + 2 \cdot \left([D_4]+ [D_4^\prime]\right) + 2 \cdot\left([D_6]+ [D_6^\prime]\right),
\eea
which leads to the D3-brane tadpole charge 
\bea
& & Q_{\rm D_3} = \frac{2}{2}+\frac{11+11+24}{6}+\frac{6\cdot 2 \cdot 11+6\cdot 2 \cdot 11+2\cdot 2 \cdot 24+2\cdot 2 \cdot 2}{48} = 24.
\eea

\item 
{\bf Model PI$_4$:} 
Four stacks of D7-branes (along with their corresponding image branes) wrapping the divisors $\{D_1, D_2, D_4, D_6\}$ are placed as follows

\bea
& & 8[O7] = \sum_{\alpha=1, 2}\, 7 \cdot \left([D_\alpha] + [D_\alpha^\prime] \right) + 1 \cdot \left([D_4]+ [D_4^\prime]\right) + 3 \cdot\left([D_6]+ [D_6^\prime]\right),
\eea
which leads to the D3-brane tadpole charge 
\bea
& & Q_{\rm D_3} = \frac{2}{2}+\frac{11+11+24}{6}+\frac{7\cdot 2 \cdot 11+7\cdot 2 \cdot 11+1\cdot 2 \cdot 24+3\cdot 2 \cdot 2}{48} = 24.
\eea

\item 
{\bf Model PI$_5$:} 
Three stacks of D7-branes (along with their corresponding image branes) wrapping the divisors $\{D_1, D_2, D_6\}$ are placed as follows

\bea
& & 8[O7] = \sum_{\alpha=1, 2}\, 8 \cdot \left([D_\alpha] + [D_\alpha^\prime] \right)  + 4 \cdot\left([D_6]+ [D_6^\prime]\right),
\eea
which leads to the D3-brane tadpole charge 
\bea
& & Q_{\rm D_3} = \frac{2}{2}+\frac{11+11+24}{6}+\frac{8\cdot 2 \cdot 11+8\cdot 2 \cdot 11+4\cdot 2 \cdot 2}{48} = 24.
\eea

\end{itemize}

\noindent
Thus we realize that the D7/O7 configurations wrap the divisors $\{D_1, D_2, D_4, D_6\}$ in all the cases, except the ``on-top" configuration which does not involve the $D_6$ divisor. Therefore, the divisor intersections of interest for inducing the KK-type and the Winding-type string loop corrections are
\bea
& & D_1 \cap D_2 = \emptyset, \qquad D_1 \cap D_4 = \emptyset, \qquad \, \, D_1 \cap D_6 = {\mathbb T}^2,\\
& & D_2 \cap D_4 = \emptyset, \qquad D_2 \cap D_6 = {\mathbb T}^2, \qquad D_4 \cap D_6 = \emptyset,\nonumber
\eea
where the respective two-cycle volumes of non-trivial intersection loci are $t_\cap(D_1 \cap D_6) = -t^1$ and $t_\cap(D_2 \cap D_6) = -t^2$. This means that Winding type string loop corrections depend only on the overall volume ${\cal V}$, and the volume of the two del-Pezzo four-cycles, and not on the inflaton modulus. However, the generic KK-type string loop corrections can arise due to the presence of non-intersecting D7/O7-planes and the O3-planes. Taking these details into account, the effective scalar potential can be read-off from the generic form presented in Eq.~(\ref{eq:Vfinal-simp}) by setting ${\cal C}_3^W = 0, \, {\cal C}_4^W = 0$. Subsequently, the simplified version of the effective scalar potential can be given as
\bea
\label{eq:Vfi-simp}
& & \hskip-0.5cm V \simeq V_0(\tau_1,\tau_2,{\cal V}, \rho_1, \rho_2) + V_{\rm PI}(\tau_4, \rho_4).
\eea
Here, the leading piece $V_0$ stabilizes all the saxionic moduli (except $\tau_4$) in the LVS minimum while the effective single field potential $V_{\rm PI}$ depends on the K3 fibre modulus $\tau_4$ as below
\bea
& & \hskip-1.3cm V_{\rm PI}(\tau_4,\rho_4) \simeq \frac{\kappa\,g_s^2\,|W_0|^2}{{\cal V}^2} \left[\frac{\gamma_b\, \left({\cal C}_3^{\rm KK}\right)^2\, \tau_4}{2{\cal V}^2} + \frac{\left({\cal C}_4^{\rm KK}\right)^2}{4{\tau_4^2}} \right] + \frac{{\cal C}_\lambda}{{\cal V}^3} \, \left(\frac{6}{\tau_4} + \frac{9\sqrt\tau_4}{{\cal V}}\right) \nonumber\\
& &  + \left(\tilde{\cal C}_{\rm poly}^{(1)} + \tau_4 \, \tilde{\cal C}_{\rm poly}^{(2)}\right) e^{-2\pi\tau_4}\, \cos(2\pi \rho_4)   + \dots, \nonumber
\eea
Now as we argued earlier, in order to have the conventional models of poly-instanton inflation, one would need to ensure that the higher derivative F$^4$ corrections are small, say via small value of the $\lambda$ parameter \cite{Cicoli:2023njy}. Moreover, one has to argue that KK-type string-loop effects can be controlled by the factor of string coupling $g_s$ or through their dependence on the complex structure moduli. In such circumstances, similar to the loop blow-up inflation \cite{Bansal:2024uzr}, one will have ``Loop Poly-instanton" or better to call as ``loop Wilson inflation" given that poly-instanton properties are no more utilized however the `Wilson' nature of divisor is still intact, similar to the del-Pezzo being part of the loop blow-up inflation. We plan to report on ``Loop Wilson inflation" in a future work.


\section{Summary and conclusions}
\label{sec_conclusions}
The single-field K\"ahler moduli based inflationary models proposed in the framework of LARGE volume scenarios have attracted a lot of attention since a couple of decades. These models, which are popularly known as  (loop) blow-up inflation, fibre inflation and poly-instanton inflation, are distinguished by the very specific geometric requirements on the compactifying CY threefolds and the internal geometries of the divisors and curves. In this article, we have proposed a unified framework to realize all these (known) LVS inflationary models using the same CY threefold. It turns out that the aforementioned three classes of LVS inflationary models can be realized by using different orientifolds of a single CY threefold. In this regard, we follow a two step process to address the demands/challenges:
\begin{itemize}

\item 
{\bf Step-1: Scanning the CY geometries} \\
The minimal requirement for such a CY threefold is to have a K3- or ${\mathbb T}^4$-fibration structure, two diagonal del-Pezzo divisors, and a so-called Wilson divisor to support the poly-instanton corrections. Subsequently one can investigate the local demands regarding the tadpole cancellations, the choice of brane-settings and the divisor intersections, for a given holomorphic involution.

\item 
{\bf Step-2: Finding the effective scalar potential} \\
For the second step, the main challenge is to find the suitable/appropriate and sufficient sub-leading corrections to the four-dimensional effective scalar potential, e.g. those classified in terms of string-loop effects and the ones arising from the higher derivative $\alpha^\prime$ series, such that just enough corrections are induced, and no unwanted contributions to the scalar potential are present, as those may potentially destroy the inflationary plateau before sufficient efolds are gained.

\end{itemize}

\noindent
For this purpose, we scanned all the CY geometries with $1\le h^{1,1}({\rm CY}) \le 6$ following from the KS database with triangulations collected in the AGHJN database. In particular, we looked at a detailed classification of the divisor topologies of all the CY geometries. In this regard, we have studied 101678 CY geometries and 984990 divisors for their Hodge diamond topology and other properties such as diagonality of divisors, vanishing of the second Chern numbers etc. In our scan we have found a total of 61 CY geometries (out of 101678) which can potentially support a unified framework for all the LVS inflationary models. 

However, the next level technical challenge is about the choice of holomorphic involution and the subsequent brane setting such that suitable and (just) enough sub-leading corrections to the scalar potential are induced. We illustrate this by considering two CY geometries with $h^{1,1}({\rm CY}) = 4$ which happen to be the only ones found in the range $ 1 \le h^{1,1}({\rm CY}) \le 4$. In order to fix the setup, we have analyzed the various brane-setting possibilities for a given reflection involution corresponding to flipping the sign of a single toric coordinate at a time: $\sigma_i : x_i \to - x_i$. This leads to $O3/O7$ fixed point set and flips the holomorphic $(3,0)$ form of the CY threefold. We subsequently discuss the various brane-setting possibilities of all the eight reflection involutions and discuss the possibilities when poly-instanton effects can be present or absent. For example, we need such corrections for poly-instanton inflation, however, for fibre inflation and the (loop) blow-up inflation we better prefer to prevent such effects from appearing in the scalar potential through a suitable choice of involution and brane setting itself. Similarly, for string-loop effects of KK- and Winding-type, we need certain (non-)intersecting D7/O7 stacks with intersection locus being a non-shrinkable curve, along with the presence of O3-planes. 

Having considered these aspects, we have computed the sub-leading contributions to the four-dimensional effective scalar potential in various settings. Subsequently, in order to connect our global models with the standard LVS inflationary analysis already present in the literature, we have accordingly recovered the desired standard forms of the scalar potentials. Here let us emphasize that our current aim in this work is not to address the viability aspects of these three LVS inflationary proposals and we limit ourselves only to an explicit realization of the standard inflationary potentials in some concrete global constructions. The open challenges such as string-loop effects in the blow-up inflation or the poly-instanton inflation, and the inflaton field-range bound problem in the fibre inflation remain open questions. On these lines, it would be interesting to study the generic potential (\ref{eq:Vfinal-simp}) for different phenomenological purposes and realizing a truly unified LVS inflation, which we plan to present in a future (companion) work.


\section*{Acknowledgments}
PS would like to thank Ross Altman for his very useful help with the AGHJN CY database, and Arthur Hebecker for his comments/clarifications regarding the loop blow-up inflation. PS gratefully acknowledges the {\it Mathematical Research Impact Centric Support {\rm (MATRICS)} grant} received from the {\it Anusandhan National Research Foundation {\rm (ANRF)}}, India. In addition, PS would like to thank the {\it Department of Science and Technology} (DST), India for the kind support.

PS is also thankful to the organizers of the conference ``Indian String Meeting (ISM)" jointly held at NISER+IIT-BBS, Bhubaneswar during Dec 09-14, 2025, and the organizers of the workshop ``Recent Progress in Computational String Geometry" - a BIRS-CMI event supported by {\it Banff International Research Station for Mathematical Innovation and Discovery} - held during January 26-31, 2026 at Chennai Mathematical Institute (CMI), Chennai, where parts of this work have been presented. 


\newpage
\appendix

\section{List of CY geometries with $h^{1,1}({\rm CY}) = 5$ suitable for unified LVS inflation}
\label{sec_appendix}
After scanning the 13494 CY geometries with $h^{1,1}({\rm CY}) = 5$, we have found only 14 candidates fulfilling the preliminary ingredients desired to achieve the framework of what we call as `unified LVS inflation'. Moreover, we also note that out of 14 candidate CY geometries, there are only 6 CY geometries that have a Wilson divisor of vanishing $\Pi$ relevant for the impact of the higher derivative F$^4$-corrections. These 14 CY geometries correspond to the polytope Ids: 2579, 2580, 2583, 3352, 4429, 4686, 4687, 4689, 5087, 5789, 6427 and 6430 in the AGHJN CY database of \cite{Altman:2014bfa}, and we also note that there are multiple CY geometries for polytope Ids 2580 and 4687. For the interested readers, we present the toric data along with the relevant divisor topologies for all the 14 CY geometries which can be potential candidates for unified LVS inflationary model building. 

\subsubsection*{Example 1}
\noindent 
Polytope Id.~: 2579 \qquad $(h^{2,1}, h^{1,1}) = (59, 5)$
\begin{center}
\begin{tabular}{|c|ccccccccc|}
\hline
\cellcolor[gray]{0.9}Hyp &\cellcolor[gray]{0.9} $x_1$  &\cellcolor[gray]{0.9} $x_2$  &\cellcolor[gray]{0.9} $x_3$  &\cellcolor[gray]{0.9} $x_4$  &\cellcolor[gray]{0.9} $x_5$ & \cellcolor[gray]{0.9}$x_6$  &\cellcolor[gray]{0.9} $x_7$ &\cellcolor[gray]{0.9} $x_8$  &\cellcolor[gray]{0.9} $x_9$    \\
\hline
\cellcolor[gray]{0.9} 3 & 0 & 0 & 0 & 0 & 1 & 1 & 0 & 1 & 0 \\
\cellcolor[gray]{0.9} 3 & 0 & 0 & 0 & 1 & 0 & 1 & 0 & 0 & 1 \\
\cellcolor[gray]{0.9} 6 & 0 & 0 & 1 & 1 & 1 & 2 & 1 & 0 & 0 \\
\cellcolor[gray]{0.9} 6 & 0 & 1 & 1 & 2 & 0 & 2 & 0 & 0 & 0 \\
\cellcolor[gray]{0.9} 6 & 1 & 0 & 1 & 0 & 2 & 2 & 0 & 0 & 0 \\
\hline
\cellcolor[gray]{0.9} & dP$_6$ & dP$_6$ & K3 &  &  &  & W & W & W \\
\hline
\end{tabular}
\end{center}
\be
{\rm SR} =  \{x_1 x_2, x_1 x_3, x_1 x_5, x_1 x_9, x_2 x_3, x_2 x_4, x_2 x_8, x_3 x_7, x_4 x_9, x_5 x_6 x_8, x_4 x_6 x_7 x_8, x_5 x_6 x_7 x_9\} \,. \nn
\ee

\subsubsection*{Example 2}
\noindent 
Polytope Id.~: 2580 \qquad $(h^{2,1}, h^{1,1}) = (59, 5)$
\begin{center}
\begin{tabular}{|c|ccccccccc|}
\hline
\cellcolor[gray]{0.9}Hyp & \cellcolor[gray]{0.9} $x_1$  &\cellcolor[gray]{0.9} $x_2$  &\cellcolor[gray]{0.9} $x_3$  &\cellcolor[gray]{0.9} $x_4$  &\cellcolor[gray]{0.9} $x_5$ & \cellcolor[gray]{0.9}$x_6$  &\cellcolor[gray]{0.9} $x_7$ &\cellcolor[gray]{0.9} $x_8$  &\cellcolor[gray]{0.9} $x_9$    \\
\hline
\cellcolor[gray]{0.9} 3 & 0 & 0 & 0 & 0 & 0 & 1 & 1 & 1 & 0 \\
\cellcolor[gray]{0.9} 3 & 0 & 0 & 0 & 0 & 1 & 0 & 1 & 0 & 1 \\
\cellcolor[gray]{0.9} 6 & 0 & 0 & 1 & 1 & 2 & 0 & 2 & 0 & 0 \\
\cellcolor[gray]{0.9} 6 & 0 & 1 & 1 & 0 & 1 & 1 & 2 & 0 & 0 \\
\cellcolor[gray]{0.9} 6 & 1 & 0 & 0 & 1 & 1 & 1 & 2 & 0 & 0 \\
\hline
\cellcolor[gray]{0.9}  & dP$_6$ & dP$_6$ & NdP$_{12}$ & NdP$_{12}$ &  & K3 &  & W & W$_\Pi$\\
\hline
\end{tabular}
\end{center}
\be
{\rm SR} =  \{x_1 x_2, x_1 x_4, x_1 x_6, x_1 x_9, x_2 x_3, x_2 x_6, x_2 x_9, x_3 x_4, x_5 x_9, x_6 x_7 x_8, x_3 x_5 x_7 x_8, x_4 x_5 x_7 x_8\} \,. \nn
\ee

\subsubsection*{Example 3}
\noindent 
Polytope Id.~: 2580 \qquad $(h^{2,1}, h^{1,1}) = (59, 5)$
\begin{center}
\begin{tabular}{|c|ccccccccc|}
\hline
\cellcolor[gray]{0.9}Hyp &\cellcolor[gray]{0.9} $x_1$  &\cellcolor[gray]{0.9} $x_2$  &\cellcolor[gray]{0.9} $x_3$  &\cellcolor[gray]{0.9} $x_4$  &\cellcolor[gray]{0.9} $x_5$ & \cellcolor[gray]{0.9}$x_6$  &\cellcolor[gray]{0.9} $x_7$ &\cellcolor[gray]{0.9} $x_8$  &\cellcolor[gray]{0.9} $x_9$    \\
\hline
\cellcolor[gray]{0.9} 3 & 0 & 0 & 0 & 0 & 0 & 1 & 1 & 1 & 0 \\
\cellcolor[gray]{0.9} 3 & 0 & 0 & 0 & 0 & 1 & 0 & 1 & 0 & 1 \\
\cellcolor[gray]{0.9} 6 & 0 & 0 & 1 & 1 & 2 & 0 & 2 & 0 & 0 \\
\cellcolor[gray]{0.9} 6 & 0 & 1 & 1 & 0 & 1 & 1 & 2 & 0 & 0 \\
\cellcolor[gray]{0.9} 6 & 1 & 0 & 0 & 1 & 1 & 1 & 2 & 0 & 0 \\
\hline
\cellcolor[gray]{0.9}  & NdP$_{12}$ & NdP$_{12}$ & dP$_6$ & dP$_6$ & K3 &  &  & W$_\Pi$ & W \\
\hline
\end{tabular}
\end{center}
\be
{\rm SR} =  \{x_1 x_2, x_1 x_4, x_2 x_3, x_3 x_4, x_3 x_5, x_3 x_8, x_4 x_5, x_4 x_8, x_6 x_8, x_5 x_7 x_9, x_1 x_6 x_7 x_9, x_2 x_6 x_7 x_9\} \,. \nn
\ee

\subsubsection*{Example 4}
\noindent 
Polytope Id.~: 2583 \qquad $(h^{2,1}, h^{1,1}) = (59, 5)$
\begin{center}
\begin{tabular}{|c|ccccccccc|}
\hline
\cellcolor[gray]{0.9}Hyp &\cellcolor[gray]{0.9} $x_1$  &\cellcolor[gray]{0.9} $x_2$  &\cellcolor[gray]{0.9} $x_3$  &\cellcolor[gray]{0.9} $x_4$  &\cellcolor[gray]{0.9} $x_5$ & \cellcolor[gray]{0.9}$x_6$  &\cellcolor[gray]{0.9} $x_7$ &\cellcolor[gray]{0.9} $x_8$  &\cellcolor[gray]{0.9} $x_9$    \\
\hline
\cellcolor[gray]{0.9} 3 & 0 & 0 & 0 & 0 & 0 & 0 & 1 & 1 & 1 \\
\cellcolor[gray]{0.9} 3 & 0 & 0 & 0 & 1 & 1 & 0 & 0 & 1 & 0 \\
\cellcolor[gray]{0.9} 3 & 0 & 0 & 1 & 0 & 0 & 1 & 0 & 1 & 0 \\
\cellcolor[gray]{0.9} 6 & 0 & 1 & 1 & 0 & 1 & 0 & 1 & 2 & 0 \\
\cellcolor[gray]{0.9} 6 & 1 & 0 & 0 & 1 & 0 & 1 & 1 & 2 & 0 \\
\hline
\cellcolor[gray]{0.9}  & dP$_6$ & dP$_6$ & NdP$_{12}$ & NdP$_{12}$ & NdP$_{12}$ & NdP$_{12}$ & K3 &  & W \\
\hline
\end{tabular}
\end{center}
\be
{\rm SR} =  \{x_1 x_2, x_1 x_4, x_1 x_6, x_1 x_7, x_2 x_3, x_2 x_5, x_2 x_7, x_3 x_6, x_4 x_5, x_7 x_8 x_9, x_3 x_5 x_8 x_9, x_4 x_6 x_8 x_9\} \,. \nn
\ee

\subsubsection*{Example 5}
\noindent 
Polytope Id.~: 3352 \qquad $(h^{2,1}, h^{1,1}) = (65, 5)$
\begin{center}
\begin{tabular}{|c|ccccccccc|}
\hline
\cellcolor[gray]{0.9}Hyp &\cellcolor[gray]{0.9} $x_1$  &\cellcolor[gray]{0.9} $x_2$  &\cellcolor[gray]{0.9} $x_3$  &\cellcolor[gray]{0.9} $x_4$  &\cellcolor[gray]{0.9} $x_5$ & \cellcolor[gray]{0.9}$x_6$  &\cellcolor[gray]{0.9} $x_7$ &\cellcolor[gray]{0.9} $x_8$  &\cellcolor[gray]{0.9} $x_9$    \\
\hline
\cellcolor[gray]{0.9} 3 & 0 & 0 & 0 & 0 & 1 & 0 & 1 & 1 & 0 \\
\cellcolor[gray]{0.9} 4 & 0 & 0 & 0 & 0 & 1 & 1 & 1 & 0 & 1 \\
\cellcolor[gray]{0.9} 7 & 0 & 1 & 1 & 0 & 2 & 1 & 2 & 0 & 0 \\
\cellcolor[gray]{0.9} 7 & 1 & 0 & 0 & 1 & 2 & 1 & 2 & 0 & 0 \\
\cellcolor[gray]{0.9} 6 & 1 & 1 & 0 & 0 & 2 & 0 & 2 & 0 & 0 \\
\hline
\cellcolor[gray]{0.9}  & NdP$_{12}$ & NdP$_{12}$ & dP$_6$ & dP$_6$ &  & K3 &  & dP$_1$ & W \\
\hline
\end{tabular}
\end{center}
\be
{\rm SR} =  \{x_1 x_2, x_1 x_4, x_2 x_3, x_3 x_4, x_3 x_6, x_3 x_8, x_4 x_6, x_4 x_8, x_6 x_9, x_5 x_7 x_8, x_1 x_5 x_7 x_9, x_2 x_5 x_7 x_9\} \,. \nn
\ee

\subsubsection*{Example 6}
\noindent 
Polytope Id.~: 4429 \qquad $(h^{2,1}, h^{1,1}) = (77, 5)$
\begin{center}
\begin{tabular}{|c|ccccccccc|}
\hline
\cellcolor[gray]{0.9}Hyp &\cellcolor[gray]{0.9} $x_1$  &\cellcolor[gray]{0.9} $x_2$  &\cellcolor[gray]{0.9} $x_3$  &\cellcolor[gray]{0.9} $x_4$  &\cellcolor[gray]{0.9} $x_5$ & \cellcolor[gray]{0.9}$x_6$  &\cellcolor[gray]{0.9} $x_7$ &\cellcolor[gray]{0.9} $x_8$  &\cellcolor[gray]{0.9} $x_9$    \\
\hline
\cellcolor[gray]{0.9} 2 & 0 & 0 & 0 & 0 & 0 & 0 & 1 & 1 & 0 \\
\cellcolor[gray]{0.9} 2 & 0 & 0 & 0 & 0 & 1 & 0 & 0 & 0 & 1 \\
\cellcolor[gray]{0.9} 3 & 0 & 0 & 0 & 0 & 1 & 1 & 0 & 1 & 0 \\
\cellcolor[gray]{0.9} 3 & 0 & 0 & 1 & 0 & 1 & 0 & 1 & 0 & 0 \\
\cellcolor[gray]{0.9} 6 & 1 & 1 & 0 & 1 & 0 & 0 & 3 & 0 & 0 \\
\hline
\cellcolor[gray]{0.9}  &  &  & ${\mathbb P}^2$ &  & K3 & NdP$_{18}$ &  & ${\mathbb P}^2$ & W$_\Pi$ \\
\hline
\end{tabular}
\end{center}
\be
{\rm SR} =  \{x_3 x_5, x_3 x_6, x_3 x_7, x_5 x_6, x_5 x_9, x_6 x_8, x_7 x_8, x_7 x_9, x_8 x_9, x_1 x_2 x_4\} \,. \nn
\ee

\subsubsection*{Example 7}
\noindent 
Polytope Id.~: 4686 \qquad $(h^{2,1}, h^{1,1}) = (81, 5)$
\begin{center}
\begin{tabular}{|c|ccccccccc|}
\hline
\cellcolor[gray]{0.9}Hyp &\cellcolor[gray]{0.9} $x_1$  &\cellcolor[gray]{0.9} $x_2$  &\cellcolor[gray]{0.9} $x_3$  &\cellcolor[gray]{0.9} $x_4$  &\cellcolor[gray]{0.9} $x_5$ & \cellcolor[gray]{0.9}$x_6$  &\cellcolor[gray]{0.9} $x_7$ &\cellcolor[gray]{0.9} $x_8$  &\cellcolor[gray]{0.9} $x_9$    \\
\hline
\cellcolor[gray]{0.9} 4 & 0 & 0 & 0 & 0 & 1 & 2 & 0 & 1 & 0 \\
\cellcolor[gray]{0.9} 4 & 0 & 0 & 0 & 1 & 0 & 2 & 0 & 0 & 1 \\
\cellcolor[gray]{0.9} 8 & 0 & 0 & 1 & 1 & 1 & 4 & 1 & 0 & 0 \\
\cellcolor[gray]{0.9} 8 & 0 & 1 & 1 & 2 & 0 & 4 & 0 & 0 & 0 \\
\cellcolor[gray]{0.9} 8 & 1 & 0 & 1 & 0 & 2 & 4 & 0 & 0 & 0 \\
\hline
\cellcolor[gray]{0.9}  & dP$_7$ & dP$_7$ & K3 &  &  &  & W & W & W \\
\hline
\end{tabular}
\end{center}
\be
{\rm SR} =  \{x_1 x_2, x_1 x_3, x_1 x_5, x_1 x_9, x_2 x_3, x_2 x_4, x_2 x_8, x_3 x_7, x_5 x_8, x_4 x_6 x_9, x_4 x_6 x_7 x_8, x_5 x_6 x_7 x_9\} \,. \nn
\ee

\subsubsection*{Example 8}
\noindent 
Polytope Id.~: 4687 \qquad $(h^{2,1}, h^{1,1}) = (81, 5)$
\begin{center}
\begin{tabular}{|c|ccccccccc|}
\hline
\cellcolor[gray]{0.9}Hyp &\cellcolor[gray]{0.9} $x_1$  &\cellcolor[gray]{0.9} $x_2$  &\cellcolor[gray]{0.9} $x_3$  &\cellcolor[gray]{0.9} $x_4$  &\cellcolor[gray]{0.9} $x_5$ & \cellcolor[gray]{0.9}$x_6$  &\cellcolor[gray]{0.9} $x_7$ &\cellcolor[gray]{0.9} $x_8$  &\cellcolor[gray]{0.9} $x_9$    \\
\hline
\cellcolor[gray]{0.9} 4 & 0 & 0 & 0 & 0 & 0 & 1 & 2 & 1 & 0 \\
\cellcolor[gray]{0.9} 4 & 0 & 0 & 0 & 0 & 1 & 0 & 2 & 0 & 1 \\
\cellcolor[gray]{0.9} 8 & 0 & 0 & 1 & 1 & 2 & 0 & 4 & 0 & 0 \\
\cellcolor[gray]{0.9} 8 & 0 & 1 & 1 & 0 & 1 & 1 & 4 & 0 & 0 \\
\cellcolor[gray]{0.9} 8 & 1 & 0 & 0 & 1 & 1 & 1 & 4 & 0 & 0 \\
\hline
\cellcolor[gray]{0.9}  & dP$_7$ & dP$_7$ & NdP$_{11}$ & NdP$_{11}$ &  & K3 &  & W & W$_\Pi$\\
\hline
\end{tabular}
\end{center}
\be
{\rm SR} =  \{x_1 x_2, x_1 x_4, x_1 x_6, x_1 x_9, x_2 x_3, x_2 x_6, x_2 x_9, x_3 x_4, x_6 x_8, x_5 x_7 x_9, x_3 x_5 x_7 x_8, x_4 x_5 x_7 x_8\}. \nn
\ee

\subsubsection*{Example 9}
\noindent 
Polytope Id.~: 4687 \qquad $(h^{2,1}, h^{1,1}) = (81, 5)$
\begin{center}
\begin{tabular}{|c|ccccccccc|}
\hline
\cellcolor[gray]{0.9}Hyp &\cellcolor[gray]{0.9} $x_1$  &\cellcolor[gray]{0.9} $x_2$  &\cellcolor[gray]{0.9} $x_3$  &\cellcolor[gray]{0.9} $x_4$  &\cellcolor[gray]{0.9} $x_5$ & \cellcolor[gray]{0.9}$x_6$  &\cellcolor[gray]{0.9} $x_7$ &\cellcolor[gray]{0.9} $x_8$  &\cellcolor[gray]{0.9} $x_9$    \\
\hline
\cellcolor[gray]{0.9} 4 & 0 & 0 & 0 & 0 & 0 & 1 & 2 & 1 & 0 \\
\cellcolor[gray]{0.9} 4 & 0 & 0 & 0 & 0 & 1 & 0 & 2 & 0 & 1 \\
\cellcolor[gray]{0.9} 8 & 0 & 0 & 1 & 1 & 2 & 0 & 4 & 0 & 0 \\
\cellcolor[gray]{0.9} 8 & 0 & 1 & 1 & 0 & 1 & 1 & 4 & 0 & 0 \\
\cellcolor[gray]{0.9} 8 & 1 & 0 & 0 & 1 & 1 & 1 & 4 & 0 & 0 \\
\hline
\cellcolor[gray]{0.9}  & NdP$_{11}$ & NdP$_{11}$ & dP$_7$ & dP$_7$ & K3 &  &  & W$_\Pi$ & W \\
\hline
\end{tabular}
\end{center}
\be
{\rm SR} =  \{x_1 x_2, x_1 x_4, x_2 x_3, x_3 x_4, x_3 x_5, x_3 x_8, x_4 x_5, x_4 x_8, x_6 x_8, x_5 x_7 x_9, x_1 x_6 x_7 x_9, x_2 x_6 x_7 x_9\} \,. \nn
\ee

\subsubsection*{Example 10}
\noindent 
Polytope Id.~: 4689 \qquad $(h^{2,1}, h^{1,1}) = (81, 5)$
\begin{center}
\begin{tabular}{|c|ccccccccc|}
\hline
\cellcolor[gray]{0.9}Hyp &\cellcolor[gray]{0.9} $x_1$  &\cellcolor[gray]{0.9} $x_2$  &\cellcolor[gray]{0.9} $x_3$  &\cellcolor[gray]{0.9} $x_4$  &\cellcolor[gray]{0.9} $x_5$ & \cellcolor[gray]{0.9}$x_6$  &\cellcolor[gray]{0.9} $x_7$ &\cellcolor[gray]{0.9} $x_8$  &\cellcolor[gray]{0.9} $x_9$    \\
\hline
\cellcolor[gray]{0.9} 4 & 0 & 0 & 0 & 0 & 0 & 0 & 1 & 2 & 1 \\
\cellcolor[gray]{0.9} 4 & 0 & 0 & 0 & 1 & 1 & 0 & 0 & 2 & 0 \\
\cellcolor[gray]{0.9} 4 & 0 & 0 & 1 & 0 & 0 & 1 & 0 & 2 & 0 \\
\cellcolor[gray]{0.9} 8 & 0 & 1 & 1 & 0 & 1 & 0 & 1 & 4 & 0 \\
\cellcolor[gray]{0.9} 8 & 1 & 0 & 0 & 1 & 0 & 1 & 1 & 4 & 0 \\
\hline
\cellcolor[gray]{0.9}  & dP$_7$ & dP$_7$ & NdP$_{11}$ & NdP$_{11}$ & NdP$_{11}$ & NdP$_{11}$ & K3 &  & W \\
\hline
\end{tabular}
\end{center}
\be
{\rm SR} =  \{x_1 x_2, x_1 x_4, x_1 x_6, x_1 x_7, x_2 x_3, x_2 x_5, x_2 x_7, x_3 x_6, x_4 x_5, x_7 x_8 x_9, x_3 x_5 x_8 x_9, x_4 x_6 x_8 x_9\} \,. \nn
\ee

\subsubsection*{Example 11}
\noindent 
Polytope Id.~: 5087 \qquad $(h^{2,1}, h^{1,1}) = (85, 5)$
\begin{center}
\begin{tabular}{|c|ccccccccc|}
\hline
\cellcolor[gray]{0.9}Hyp &\cellcolor[gray]{0.9} $x_1$  &\cellcolor[gray]{0.9} $x_2$  &\cellcolor[gray]{0.9} $x_3$  &\cellcolor[gray]{0.9} $x_4$  &\cellcolor[gray]{0.9} $x_5$ & \cellcolor[gray]{0.9}$x_6$  &\cellcolor[gray]{0.9} $x_7$ &\cellcolor[gray]{0.9} $x_8$  &\cellcolor[gray]{0.9} $x_9$    \\
\hline
\cellcolor[gray]{0.9} 4 & 0 & 0 & 0 & 0 & 0 & 0 & 1 & 2 & 1 \\
\cellcolor[gray]{0.9} 4 & 0 & 0 & 0 & 0 & 1 & 1 & 0 & 2 & 0 \\
\cellcolor[gray]{0.9} 8 & 0 & 0 & 1 & 1 & 0 & 1 & 1 & 4 & 0 \\
\cellcolor[gray]{0.9} 8 & 0 & 1 & 1 & 0 & 0 & 0 & 2 & 4 & 0 \\
\cellcolor[gray]{0.9} 10 & 1 & 0 & 1 & 0 & 0 & 2 & 1 & 5 & 0 \\
\hline
\cellcolor[gray]{0.9}  & NdP$_{11}$ & NdP$_{10}$ &  & ${\mathbb P}^1\times{\mathbb P}^1$ & ${\mathbb P}^1\times{\mathbb P}^1$ & K3 &  &  & W$_\Pi$\\
\hline
\end{tabular}
\end{center}
\be
{\rm SR} =  \{x_1 x_6, x_2 x_3, x_3 x_4, x_4 x_6, x_4 x_9, x_5 x_6, x_1 x_2 x_7, x_1 x_3 x_7, x_1 x_7 x_9,  x_2 x_5 x_8, x_4 x_5 x_8, x_5 x_8 x_9, x_7 x_8 x_9\} \,. \nn
\ee

\subsubsection*{Example 12}
\noindent 
Polytope Id.~: 5789 \qquad $(h^{2,1}, h^{1,1}) = (97, 5)$
\begin{center}
\begin{tabular}{|c|ccccccccc|}
\hline
\cellcolor[gray]{0.9}Hyp &\cellcolor[gray]{0.9} $x_1$  &\cellcolor[gray]{0.9} $x_2$  &\cellcolor[gray]{0.9} $x_3$  &\cellcolor[gray]{0.9} $x_4$  &\cellcolor[gray]{0.9} $x_5$ & \cellcolor[gray]{0.9}$x_6$  &\cellcolor[gray]{0.9} $x_7$ &\cellcolor[gray]{0.9} $x_8$  &\cellcolor[gray]{0.9} $x_9$    \\
\hline
\cellcolor[gray]{0.9} 4 & 0 & 0 & 0 & 0 & 0 & 1 & 2 & 1 & 0 \\
\cellcolor[gray]{0.9} 6 & 0 & 0 & 0 & 0 & 1 & 1 & 3 & 0 & 1 \\
\cellcolor[gray]{0.9} 10 & 0 & 1 & 1 & 0 & 1 & 2 & 5 & 0 & 0 \\
\cellcolor[gray]{0.9} 10 & 1 & 0 & 0 & 1 & 1 & 2 & 5 & 0 & 0 \\
\cellcolor[gray]{0.9} 8 & 1 & 1 & 0 & 0 & 0 & 2 & 4 & 0 & 0 \\
\hline
\cellcolor[gray]{0.9}  & NdP$_{11}$ & NdP$_{11}$ & dP$_7$ & dP$_7$ & K3 &  &  & dP$_1$ & W \\
\hline
\end{tabular}
\end{center}
\be
{\rm SR} =  \{x_1 x_2, x_1 x_4, x_2 x_3, x_3 x_4, x_3 x_5, x_3 x_8, x_4 x_5, x_4 x_8, x_6 x_8, x_5 x_7 x_9, x_1 x_6 x_7 x_9, x_2 x_6 x_7 x_9\} \,. \nn
\ee

\subsubsection*{Example 13}
\noindent 
Polytope Id.~: 6427 \qquad $(h^{2,1}, h^{1,1}) = (185, 5)$
\begin{center}
\begin{tabular}{|c|ccccccccc|}
\hline
\cellcolor[gray]{0.9}Hyp &\cellcolor[gray]{0.9} $x_1$  &\cellcolor[gray]{0.9} $x_2$  &\cellcolor[gray]{0.9} $x_3$  &\cellcolor[gray]{0.9} $x_4$  &\cellcolor[gray]{0.9} $x_5$ & \cellcolor[gray]{0.9}$x_6$  &\cellcolor[gray]{0.9} $x_7$ &\cellcolor[gray]{0.9} $x_8$  &\cellcolor[gray]{0.9} $x_9$    \\
\hline
\cellcolor[gray]{0.9} 6 & 0 & 0 & 0 & 0 & 2 & 3 & 0 & 0 & 1 \\
\cellcolor[gray]{0.9} 12 & 0 & 0 & 1 & 0 & 4 & 6 & 0 & 1 & 0 \\
\cellcolor[gray]{0.9} 18 & 0 & 0 & 1 & 1 & 6 & 9 & 1 & 0 & 0 \\
\cellcolor[gray]{0.9} 12 & 0 & 1 & 0 & 1 & 4 & 6 & 0 & 0 & 0 \\
\cellcolor[gray]{0.9} 24 & 1 & 0 & 2 & 1 & 8 & 12 & 0 & 0 & 0 \\
\hline
\cellcolor[gray]{0.9}  & dP$_8$ & dP$_8$ &  & K3 &  &  & W & W & dP$_1$ \\
\hline
\end{tabular}
\end{center}
\be
{\rm SR} =  \{x_1 x_2, x_1 x_3, x_1 x_4, x_1 x_9, x_2 x_4, x_2 x_8, x_2 x_9, x_3 x_8, x_4 x_7, x_5 x_6 x_9, x_3 x_5 x_6 x_7, x_5 x_6 x_7 x_8\} \,. \nn
\ee

\subsubsection*{Example 14}
\noindent 
Polytope Id.~: 6430 \qquad $(h^{2,1}, h^{1,1}) = (185, 5)$
\begin{center}
\begin{tabular}{|c|ccccccccc|}
\hline
\cellcolor[gray]{0.9}Hyp &\cellcolor[gray]{0.9} $x_1$  &\cellcolor[gray]{0.9} $x_2$  &\cellcolor[gray]{0.9} $x_3$  &\cellcolor[gray]{0.9} $x_4$  &\cellcolor[gray]{0.9} $x_5$ & \cellcolor[gray]{0.9}$x_6$  &\cellcolor[gray]{0.9} $x_7$ &\cellcolor[gray]{0.9} $x_8$  &\cellcolor[gray]{0.9} $x_9$    \\
\hline
\cellcolor[gray]{0.9}  6 & 0 & 0 & 0 & 0 & 0 & 2 & 3 & 0 & 1 \\
\cellcolor[gray]{0.9}  12 & 0 & 0 & 0 & 0 & 1 & 4 & 6 & 1 & 0 \\
\cellcolor[gray]{0.9}  12 & 0 & 0 & 1 & 1 & 0 & 4 & 6 & 0 & 0 \\
\cellcolor[gray]{0.9}  18 & 0 & 1 & 1 & 0 & 1 & 6 & 9 & 0 & 0 \\
\cellcolor[gray]{0.9}  18 & 1 & 0 & 0 & 1 & 1 & 6 & 9 & 0 & 0 \\
\hline
\cellcolor[gray]{0.9}  & dP$_8$ & dP$_8$ & NdP$_{10}$ & dP$_{10}$ & K3 &  &  & W & dP$_1$ \\
\hline
\end{tabular}
\end{center}
\be
{\rm SR} =  \{x_1 x_2, x_1 x_4, x_1 x_5, x_1 x_9, x_2 x_3, x_2 x_5, x_2 x_9, x_3 x_4, x_5 x_8, x_6 x_7 x_9, x_3 x_6 x_7 x_8, x_4 x_6 x_7 x_8\} \,. \nn
\ee
\vskip0.2cm 

\noindent
Here NdP$_n$ refers to what we call as rigid-but-not-del-Pezzo due to their similar Hodge diamond:
\bea
& & {\rm NdP}_n \equiv \begin{tabular}{ccccc}
    & & 1 & & \\
   & 0 & & 0 & \\
  0 & & $n+1$ & & 0 \\
   & 0 & & 0 & \\
    & & 1 & & \\
  \end{tabular}; \qquad n \geq 9. \nonumber
\eea



\bibliographystyle{utphys}
\bibliography{reference}

\end{document}